\documentclass[10pt,twocolumn,journal]{IEEEtranTCOM}
\normalsize
\usepackage{times}
\usepackage{amsbsy}
\usepackage{amssymb}
\usepackage{fontenc}
\usepackage[usenames,dvipsnames]{color}
\usepackage{latexsym}
\usepackage{amsmath}
\usepackage[ansinew]{inputenc}
\usepackage[dvips ]{graphicx}
\input epsf
\usepackage{epic,eepic}
\usepackage{url}
\usepackage{caption}
\usepackage[caption=false,font=footnotesize]{subfig}
\newcommand{\revn}[1]{#1}
\newcommand{\revns}[1]{#1}
\newcommand{\revnss}[1]{#1}
\IEEEoverridecommandlockouts

\begin{document}
\title{\huge Multi-Branch Tomlinson-Harashima Precoding for MU-MIMO Systems: Theory and Algorithms}
\author{\IEEEauthorblockN {Keke Zu, Rodrigo C. de Lamare, {\it Senior Member, IEEE} and Martin Haardt, {\it Senior Member, IEEE}}

\thanks{Parts of this work have been published at the ITG/IEEE Workshop on Smart Antennas, Dresden, Germany, Mar. 2012 (see reference \cite{Keke}).}
}
\markboth{IEEE Transactions on Communications}%
{Submitted paper}

\maketitle\thispagestyle{empty} \vspace*{-1.5em}
% \maketitle

\begin{abstract}
Tomlinson-Harashima precoding (THP) is a nonlinear processing
technique employed at the transmit side and is a dual to the
successive interference cancelation (SIC) detection at the receive
side. Like SIC detection, the performance of THP strongly depends on
the ordering of the precoded symbols. The optimal ordering
algorithm, however, is impractical for multiuser MIMO (MU-MIMO)
systems with multiple receive antennas due to the fact that the
users are geographically distributed. In this paper, we propose a
multi-branch THP (MB-THP) scheme and algorithms that employ multiple
transmit processing and ordering strategies along with a selection
scheme to mitigate interference in MU-MIMO systems. Two types of
multi-branch THP (MB-THP) structures are proposed. The first one
employs a decentralized strategy with diagonal weighted filters at
the receivers of the users and the second uses a diagonal weighted
filter at the transmitter. The MB-MMSE-THP algorithms are also
derived based on an extended system model with the aid of an LQ
decomposition, which is much simpler compared to the conventional
MMSE-THP algorithms. Simulation results show that a better bit error
rate (BER) performance can be achieved by the proposed MB-MMSE-THP
precoder with a small computational complexity increase.
\end{abstract}


% \IEEEpeerreviewmaketitle
\section{Introduction}
\subsection{Background and Problem Formulation}
Multi-user MIMO (MU-MIMO) systems are promising for downlink
wireless transmissions since they can improve the average user
spectral efficiency \cite{LTEA}. When channel state information
(CSI) is available at the transmit side, precoding techniques can be
employed at the base station (BS) to mitigate the Multiuser
Interference (MUI). Then, the required computational effort for each
user's receiver can be reduced and eventually the receiver structure
can be simplified \cite{Tse}. For these reasons, the design of
cost-effective precoders is particularly important for the downlink
of MU-MIMO systems.

Channel inversion based linear precoding techniques such as zero
forcing (ZF) and minimum mean squared error (MMSE) precoding
\cite{Windpassinger} - \cite{Michael} are attractive due to their
simplicity. However, channel inversion based precoding techniques
require a higher average transmit power than other precoding
algorithms especially for ill conditioned channel matrices, which
could result in a reduced bit error ratio (BER) performance
\cite{Peel}. As a generalization of ZF precoding, block
diagonalization (BD) based precoding algorithms have been proposed
in \cite{Spencer,Choi} for MU-MIMO systems. However, BD based
precoding algorithms only take the MUI into account and thus suffer
a performance loss at low signal to noise ratios (SNRs) when the
noise is the dominant factor. A regularized block diagonalization
(RBD) precoding algorithm which introduces a regularization factor
to take the noise term into account has been proposed in
\cite{Veljko}. The performance is improved by RBD precoding, but the
BD-type precoding algorithms still cannot achieve the maximum
transmit diversity. A nonlinear vector perturbation (VP) approach,
which is based on sphere encoding (SE) to perturb the data, was
proposed in \cite{Hochwald}. With the perturbation, a near optimal
performance is achieved by VP precoding. However, finding the
optimal perturbation vector can be a nondeterministic polynomial
time (NP)-hard problem.

\subsection{Prior Art}

Another nonlinear and data-modifying technique is the dirty paper
coding (DPC) proposed in \cite{Costa}. It was shown that the
capacity of systems using DPC with independent and identically
distributed (i.i.d.) Gaussian interference is equal to that of
interference-free systems. However, DPC is not suitable for
practical use due to the requirement of infinitely long codewords
\cite{Khina}. Tomlinson-Harashima precoding (THP) \cite{Tomlinson,
Harashima} is a pre-equalization technique originally proposed for
channels with intersymbol interference (ISI). Then, the THP
technique was extended from temporal equalization to spatial
equalization for MIMO precoding in \cite{Fischer}. The details of
THP algorithms are illustrated in Section II. Although THP suffers a
performance loss compared to DPC as shown in \cite{WY}, it can work
as a cost-effective replacement of DPC in practice \cite{Erez}. As
reported in \cite{Fischer, Christoph}, the THP structure can be seen
as the dual of successive interference cancelation (SIC) detection
implemented at the receive side. Like SIC detection, the performance
of THP systems strongly depends on the ordering of the precoded
symbols.

A V-BLAST like ordering strategy for THP has been studied in
\cite{Windpassinger02} - \cite{Joham02}. The V-BLAST ordering
requires multiple calculations of the pseudo inverse of the channel
matrix. Therefore, a suboptimal heuristic sorted LQ decomposition
algorithm has been extended from the sorted QR decomposition in
\cite{Wubben, Wubben02} to THP and a tree search (TS) algorithm has
also been proposed in \cite{Habendorf}. Researchers in \cite{Jia01,
Jia02} noticed the importance of the ordering to the THP performance
as well, and the best-first ordering approach has been proposed to
perform the ordering. Algorithms for finding the near-optimal order
are proposed in \cite{Fung, Dao}. The above ordering algorithms,
however, assume that each distributed receiver is equipped with a
single antenna. Therefore, cooperative ordering processing is
impractical for distributed receivers with multiple antennas. In
\cite{Veljko2}, a successive optimization THP (SO-THP) algorithm has
been proposed for users with multiple antennas, but SO-THP only
offers a small BER gain over THP at low SNRs. For high SNRs, the BER
performance of SO-THP is comparable to that achieved by the
conventional THP algorithm. In order to achieve a better BER
performance in the whole SNR range, a novel THP structure is
proposed in this work based on a multiple-branch (MB) strategy for
MU-MIMO systems with multiple antennas at each receiver. Although
the MB-THP structure for single-user MIMO (SU-MIMO) systems has been
studied in \cite{Keke}, the original structure cannot be applied to
MU-MIMO systems since the users are physically distributed.

\subsection{Contributions}

In the literature, there are two basic THP structures according to
the position of the diagonal weighted filters, decentralized filters
located at the receivers or centralized filters deployed at the
transmitter, which are denoted as dTHP or cTHP, respectively
\cite{MH}. Most of the previous research works on THP, however, have
only focused on one of the structures. In this work, we develop
MB-THP techniques for both of the two basic THP structures. We
derive the MMSE precoding filters using an LQ decomposition. Then,
we present a design strategy for the transmit patterns that
implements an effective ordering of the data streams along with a
selection criterion for the best pattern. An analysis and a
comparison between MB-dTHP and MB-cTHP are also illustrated. By
utilizing the MB strategy, the transmit diversity gain is maximized
for MU-MIMO systems with spatial multiplexing. Therefore, the final
BER performance is improved by the proposed MB-THP algorithms. The
main contributions of the work can be summarized as
\begin{enumerate}
   \item Novel MB-THP algorithms are developed based on two basic THP structures.
   \item Cost-effective MMSE filters are derived based on the LQ decomposition of an extended matrix along with the design of transmit patterns and a selection procedure.
   \item A comprehensive performance analysis is carried out in terms of the error covariance matrix, the sum-rate and the computational complexity.
   \item A study of the most relevant precoding algorithms reported in the literature and the proposed MB-THP algorithm is conducted.
 \end{enumerate}

This paper is organized as follows. The system model and the basics of THP techniques are described in Section II. The proposed MB-THP scheme and algorithms are described in detail in Section III. A performance analysis of the existing and proposed precoders is developed in Section IV. Simulation results and conclusions are presented in Section V and Section VI, respectively.

\section{System Model and THP Algorithms}

We consider an uncoded MU-MIMO broadcast channel, with $N_t$ transmit antennas
at the base station (BS) and $N_k$ receive antennas at the $k$th
user equipment (UE). With $K$ users in the system, the total number
of receive antennas is $N_r=\sum _{k=1}^{K}N_k$. When $N_r=N_t$, the channel matrix is a square matrix. When $N_r\geq N_t$, a scheduling procedure is first performed to generate a square equivalent channel matrix. The total number of transmitted streams is denoted by $S$, and the channel is assumed to be always a square matrix, that is ${\boldsymbol H}=[{\boldsymbol H^T_1},{\boldsymbol H^T_2},\cdots,{\boldsymbol H^T_K}]^T\in\mathbb{C}^{S\times S}$ is the combined channel matrix and $\boldsymbol H_k\in\mathbb{C}^{N_k\times S}$ is the $k$th user's channel matrix.
Note that power-loading schemes \cite{Christoph} could be used to
determine the number of data streams or allocate more power to a
weaker user to improve the overall performance. However, for
simplicity, we assume that all data streams are active and \revn{equal power
loading between users and streams is performed since the power allocation is not the focus of this paper}.

\subsection{Two Basic THP Structures}
\begin{figure}[htp]
\begin{center}
\def\epsfsize#1#2{0.95\columnwidth}
\epsfbox{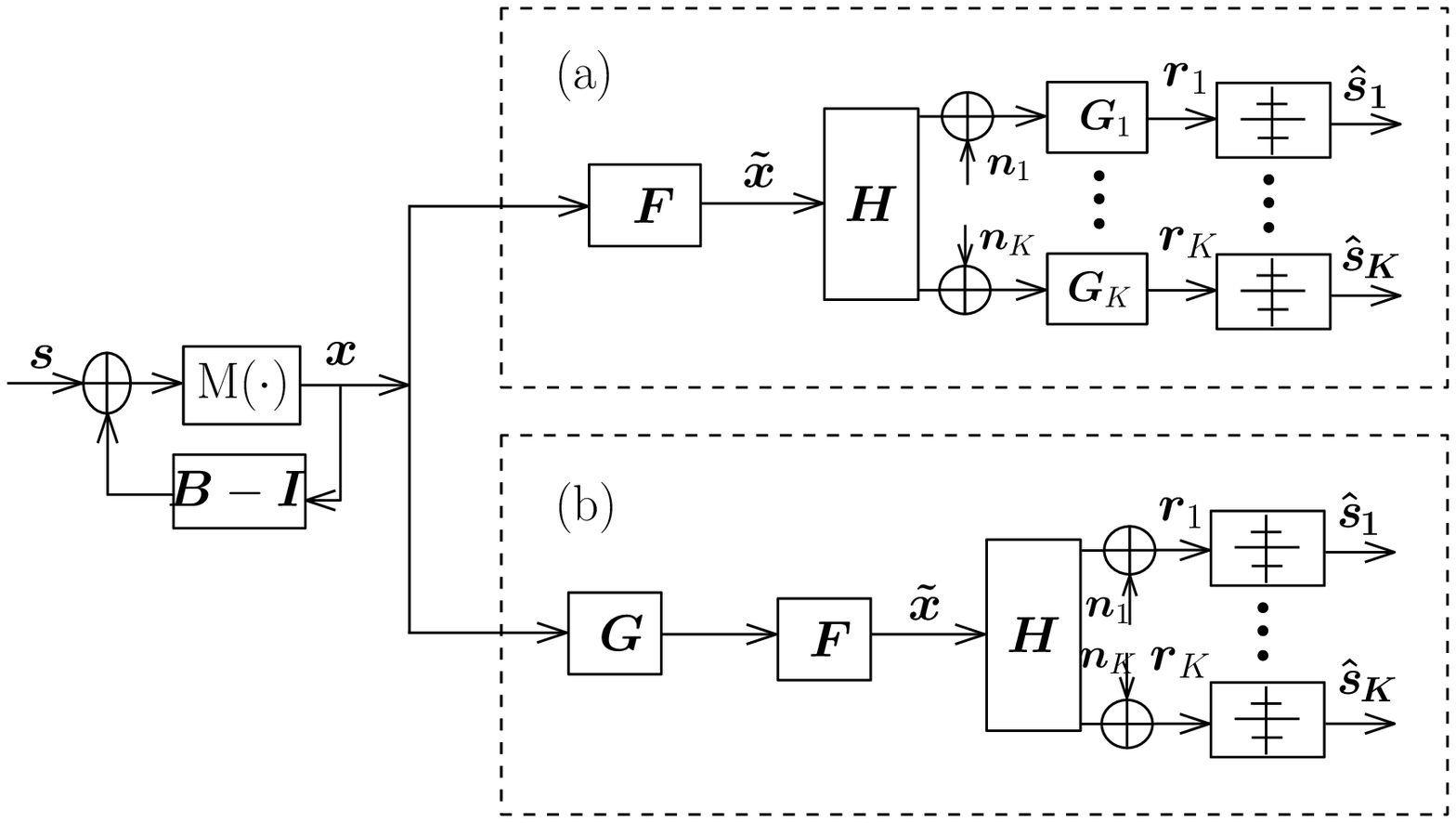} \vspace{-0.8em}
\caption{\footnotesize The two basic THP structures\\
(a) Decentralized THP: the scaling matrix $\boldsymbol G$ is separately placed at the receivers.\\
(b) Centralized THP: the scaling matrix $\boldsymbol G$ is placed at the transmitter.} \label{Two_THP_Structures}
\end{center}
\end{figure}

Based on the knowledge of CSI at the transmit side, the
interference of the parallel streams of a MIMO system with spatial
multiplexing can be subtracted from the current stream. This
SIC technique at the transmit side is known as THP and can be seen as the dual of SIC detection at the receive side.
Generally, there are three filters to implement THP algorithms: the
feedback filter $\boldsymbol B\in\mathbb{C}^{S\times S}$, the feedforward filter $\boldsymbol F\in\mathbb{C}^{S\times S}$, and the scaling matrix $\boldsymbol G\in\mathbb{C}^{S\times S}$.
According to the position of $\boldsymbol G$, there are two basic THP structures, which are illustrated in Fig. \ref {Two_THP_Structures}. The decentralized THP (dTHP) employs $\boldsymbol G$ (or sub-matrices of it) at the receivers, whereas the centralized THP (cTHP) uses $\boldsymbol G$ at the transmitter.

The feedback filter $\boldsymbol B$ is used to successively cancel the interference caused by the previous streams from the current
stream. Therefore, the feedback filter $\boldsymbol B$ should be a lower triangular matrix with ones on the main diagonal \cite{Christoph}.
The feedforward filter $\boldsymbol F$ is used to enforce the spatial causality and has to be implemented at the transmit side for MU-MIMO systems because the physically distributed users cannot be processed jointly. The scaling filter $\boldsymbol G$ contains the corresponding weighted coefficient for each stream and thus it should have a diagonal structure. The quantity $\boldsymbol x\in\mathbb{C}^{S\times 1}$ is the combined transmit signal vector after the feedback operation and $\tilde {\boldsymbol x}$ is the combined transmit signal vector after precoding, $\tilde {\boldsymbol x}=\boldsymbol F{\boldsymbol x}$ for dTHP and $\tilde {\boldsymbol x}=\boldsymbol F\boldsymbol G{\boldsymbol x}$ for cTHP.
Finally, the received signal after the feedback, feedforward, and the scaling filter, for the dTHP and cTHP is respectively given by
\begin{eqnarray}
{\boldsymbol r}^{\rm (dTHP)}=\boldsymbol G({\boldsymbol H}\boldsymbol F{\boldsymbol x} +\boldsymbol n),
\end{eqnarray}
\begin{eqnarray}
 {\boldsymbol r}^{\rm (cTHP)}=\beta({\boldsymbol H}\cdot{1\over\beta}\boldsymbol F\boldsymbol G{\boldsymbol x} +\boldsymbol n),
\end{eqnarray}
where the quantity $\boldsymbol n=[{\boldsymbol n^T_1},{\boldsymbol n^T_2},\cdots,{\boldsymbol n^T_K}]^T\in\mathbb{C}^{S\times 1} $ is the combined Gaussian noise vector with i.i.d. entries of zero mean and variance $\sigma_n^2$.
The factor $\beta$ is used to impose the power constraint ${\rm E}\| \tilde {\boldsymbol x}\|^2=\xi$ with $\xi$ being the average transmit power.

\subsection{Review of THP Algorithms}
As reported in the literature, SIC detection can be efficiently implemented by a QR decomposition \cite{Shiu}, whereas THP can be implemented by an LQ decomposition. By utilizing an LQ decomposition on the channel matrix $\boldsymbol H$, we have
\begin{eqnarray}
\boldsymbol {H}=\boldsymbol L\boldsymbol Q,
\end{eqnarray}
where $\boldsymbol L$ is a lower triangular matrix and $\boldsymbol Q$ is a unitary matrix (by unitary we mean $\boldsymbol Q^{H}\boldsymbol Q=\boldsymbol Q\boldsymbol Q^{H}=\boldsymbol I$). Therefore, the filters for the THP algorithm can be obtained as
\begin{align}
\boldsymbol F={\boldsymbol Q}^H,\\
\boldsymbol G={\rm {diag}}[l_{1,1},l_{2,2},\cdots,l_{S,S}]^{-1},\\
\boldsymbol B^{\rm (dTHP)}=\boldsymbol G\boldsymbol L, \boldsymbol B^{\rm (cTHP)}=\boldsymbol L\boldsymbol G,
\end{align}
where $l_{i,i}$ is the $i$th diagonal element of the matrix $\boldsymbol L$.

From Fig. \ref {Two_THP_Structures}, the transmitted symbols $x_i$ are successively generated as
\begin{align}
 x_i={s_i- \sum _{j=1}^{i-1}b_{i,j}{x_j}},~ {i=1,\cdots,S},
\end{align}
where $s_i$ is the $i$th transmit data with variance $\sigma_s^2$ and $b_{i,j}$ are the elements of $\boldsymbol B$ in row $i$ and column $j$.
From the above formulation, the transmit power will be significantly increased as the amplitude of $x_i$ exceeds the modulation boundary by the successive cancelation.
In order to reduce the amplitude of the channel symbol $x_i$ to the boundary of the modulation alphabet, a modulo operation ${\rm M}(\cdot)$
should be employed which is defined element-wise as \cite{Dietrich}
\begin{align}
 {\rm M}(x_i)=x_i- \biggl\lfloor {{\rm{Re}}(x_i)\over \tau }+{1\over 2}\biggr\rfloor\tau-j\biggl\lfloor {{\rm{Im}}(x_i)\over \tau }+{1\over 2}\biggr\rfloor\tau,
\end{align}
where $\tau$ is a constant for the periodic extension of the constellation. The specific value of $\tau$ depends on the chosen modulation alphabet.
Common choices for $\tau$ are $\tau=2\sqrt 2$ for QPSK symbols and $\tau=8\sqrt {10}$ in case of rectangular 16-QAM when the symbol variance is one \cite{Dietrich}.
The modulo processing is equivalent to adding a perturbation vector $\boldsymbol d$ to the transmit data $\boldsymbol s$,
such that the modified transmit data are \cite{Joham02}
\begin{align}
\boldsymbol v=\boldsymbol s +\boldsymbol d.
\end{align}
Thus, the initial signal constellation is extended periodically and the effective $k$th transmit data symbols $\boldsymbol v_k$ are taken from the expanded set.

Although the modulo operation is employed to restrict the amplitude of $\boldsymbol x$ within the same scale as that of $\boldsymbol s$,
a power loss is introduced by the nonlinear processing of THP, which can be measured by $\alpha={M\over M-1}$ for the M-QAM constellations \cite{WY, Christoph}.
The power loss is not negligible for small modulation sizes, but for moderate sizes of $M$ it is negligible and vanishes as $M$ increases.
Except for the power loss, a modulo loss is also introduced by THP due to the received symbols at the boundary of a constellation may be mistaken for symbols at the opposite boundary \cite{WY}. The modulo loss is more significant for the small constellations. We neglect the power and modulo loss in this work since moderate sizes of $M$ are employed. Then, we have ${\rm E}\| {\boldsymbol x}\|\approx{\rm E}\| {\boldsymbol s}\|$. Since the statistical property of $\boldsymbol x$ is not changed by the multiplication of the unitary matrix $\boldsymbol F$, the normalization factor $\beta$ is not necessary for dTHP. For cTHP, since the power and modulo loss can be neglected, the normalization factor is approximately obtained as
\begin{align}
\beta={{\rm E}\| {\boldsymbol F\boldsymbol G\boldsymbol x}\|\over{\rm E}\| {\boldsymbol s}\|}\approx \sqrt{\sum_{i=1}^{S}(1/l_{i,i}^2)}.
\end{align}

Mathematically, the feedback processing is equivalent to an inversion operation
${\boldsymbol B}^{-1}$. Therefore, the transmitted symbol $\boldsymbol x$ can be written as
\begin{align}
 \boldsymbol x={\boldsymbol B}^{-1}\boldsymbol v={\boldsymbol B}^{-1}(\boldsymbol s +\boldsymbol d),
\end{align}
Then, the received signal for dTHP and cTHP can be respectively expressed as
\begin{eqnarray}
{{\boldsymbol r}^{\rm (dTHP)}}&=&\boldsymbol v +\boldsymbol G\boldsymbol n,\\
 {{\boldsymbol r}^{\rm (cTHP)}}&=&\boldsymbol v +\beta\boldsymbol n.
\end{eqnarray}

\section{Proposed MB-THP Precoding Algorithm}

In this section, we first analyze the interference of the two basic
THP structures and show that the ordering of the precoded symbols
plays an important role for both of them. Based on this analysis,
the structure of the MB-cTHP and MB-dTHP precoding techniques are
proposed and illustrated. Especially for the MU-MIMO setting with
multiple receive antennas a cost-effective transmit pattern is
developed, and a selection criterion is also deduced for both of the
MB-cTHP and MB-dTHP algorithms. Finally, since the MMSE-THP
structures are the main focus of this paper, filters for
MB-MMSE-cTHP and MB-MMSE-dTHP are derived based on an extended
system model which is much simpler from a computational point of
view, as compared to conventional MMSE-THP techniques reported in
the literature so far.

\subsection{Motivation of the Proposed MB-THP Algorithm}%{Interference Analysis}

As shown in equations (12) and (13), the MU-MIMO channel is
decomposed into parallel additive white Gaussian noise (AWGN)
channels by the successive THP processing. With the power and modulo
loss ignored, the power of $\boldsymbol v$ is approximately equal to
that of $\boldsymbol s$. Then, the error covariance matrices of the
effective transmit signal $\boldsymbol v$ for dTHP and cTHP schemes
are respectively given by
\begin{eqnarray}
{\boldsymbol \Phi}_{\rm {dTHP}}&=&{\rm diag}(\sigma_n^2/{l^2_{1,1}},\cdots,\sigma_n^2/{l^2_{S,S}}),\\
{\boldsymbol \Phi}_{\rm {cTHP}}&=&{\rm diag}(\sigma_n^2 \sum_{i=1}^{S}(1/l_{i,i}^2),\cdots,\sigma_n^2 \sum_{i=1}^{S}(1/l_{i,i}^2)).
\end{eqnarray}

From (14) and (15), we can verify that the error covariance matrices
are different among layers for dTHP while they are equal for cTHP.
Therefore, for each layer, the SNR is inversely proportional to
$1/l_{i,i}^2$ for dTHP, while it is inversely proportional to
$\sum_{i=1}^{S}(1/l_{i,i}^2)$ for cTHP. Due to the lower triangular
structure of the feedback matrix $\boldsymbol B$, the interference
from the transmitted data $s_1, s_2,\cdots,s_S$ is canceled out from
$s_1$ to $s_S$ in dTHP. That is, the layer precoded first will
interfere with the layer precoded afterward. Then, the performance
of dTHP will be dominated by the layer with the minimum SNR. For
cTHP, the sum $\sum_{i=1}^{S}(1/l_{i,i}^2)$ can be influenced by
reordering the rows of $\boldsymbol H$ during the LQ decomposition.
It is worth noting that the row pivoting known in mathematics is
used for the LQ decomposition when a matrix $\boldsymbol B$ is rank
deficient, that is
\begin{align}
\boldsymbol P\boldsymbol B=\boldsymbol L\boldsymbol Q,
\end{align}
where the row permutation matrix $\boldsymbol P$ is chosen so that the diagonal elements of $|\boldsymbol L|$ are decreasing with $|\cdot|$ being the element-wise absolute value operation. The specific requirement of the row permutation matrix $\boldsymbol P$ does not take the physical location of the receive antennas into account, which prohibits the application of the row pivoting scheme in MU-MIMO systems with multiple receive antennas since the data streams that belong to one user may be allocated to other users. For the special case when all distributed users are equipped with a single antenna, the row permutation matrix $\boldsymbol P$ needs to be calculated for each transmission when the channel changes to ensure a decreasing order.

In particular, the ordering of the precoded symbols plays an
important role in the performance of THP systems. Thus, considerable
research efforts have been spent on the development of various
ordering methods \cite{Windpassinger02} - \cite{Dao}. However, they
all focused on SU-MIMO or MU-MIMO systems with single receive
antenna. For MU-MIMO systems with multiple receive antennas, these
cooperative ordering algorithms are impractical due to the
geographically distributed users. In addition, most of the ordering
algorithms only consider one THP structure, either cTHP or dTHP.

In this work, a MB-THP structure with efficient transmit pattern
design, which is predefined and especially suited for the users
equipped with multiple antennas, is proposed based on the two basic
THP structures. The design of transmit patterns is performed in two
steps. In the first step we get the ordering patterns $\boldsymbol
T_u^{(i)}~{\rm for}~i=1,\cdots,K$ between multiple users. In the
second step, we obtain the ordering patterns ${\boldsymbol
T_{k_i}^{(j)}}$ between multiple streams for the $i$th user with $j$
denoting the different ordering states.

\subsection{Structure of the Proposed MB-THP}

The idea of multi-branch (MB) processing has been first proposed in
\cite{Rodrigo} as the parallel arbitrated branches to improve the
performance of decision feedback (DF) receivers. MB-SIC detectors
have been proposed in \cite{Rui,mbdf} to exploit diversity gains in
MIMO systems. In \cite{Rodrigo02}, the authors applied the MB
strategy to generate interleaving patterns for DS-CDMA systems.
Inspired by these research works, the MB-THP algorithms for the
MU-MIMO downlink are developed and proposed in this work. The
structures of the proposed MB-THP schemes are illustrated in Fig. 2.
\begin{figure*}[htp]
\begin{center}
\def\epsfsize#1#2{1.5\columnwidth}
\epsfbox{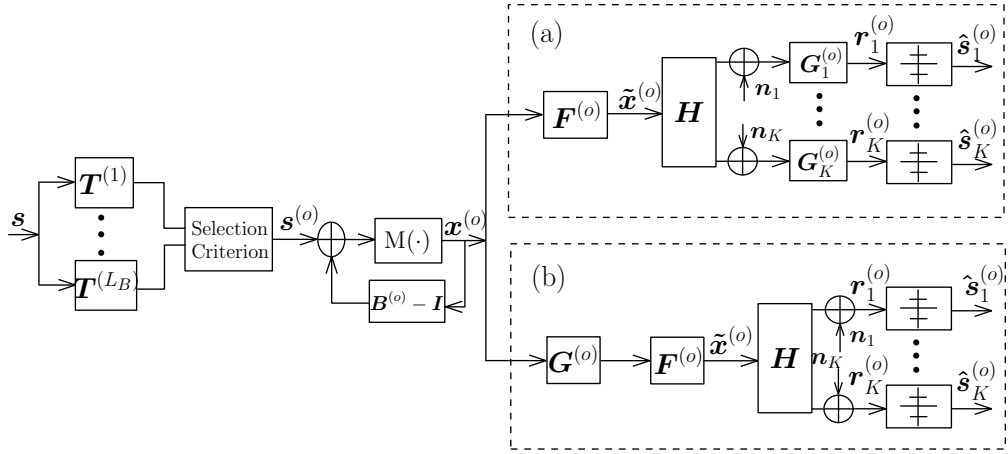} \\%\vspace{-0.8em}\\
\caption{\footnotesize The proposed MB-THP structures (a) MB-dTHP (b) MB-cTHP}
\end{center}
\end{figure*}
The matrices $\boldsymbol T^{(l)}\in\mathbb{C}^{N_r\times
N_r}~(l=1,\cdots,L_B)$ are the transmit patterns used to generate
multiple parallel candidate branches, where $L_B$ is the total
number of branches. A proper selection metric is employed to choose
the optimal branch to transmit the data streams. Then, the matrices
$\boldsymbol B^{(o)}$, $\boldsymbol F^{(o)}$ and $\boldsymbol
G^{(o)}$ represent the feedback, feedforward and scaling filters for
the selected branch.

\subsection{Design of the Transmit Patterns}

One of the objectives of this work is to design transmit patterns
that are effective and simple. Observing the formulation in (14) and
(15), the SNR performance of dTHP and cTHP can be influenced by
${\boldsymbol \Phi}_{\rm {dTHP}}$ and ${\boldsymbol \Phi}_{\rm
{cTHP}}$. An ordering of the rows of $\boldsymbol H$ will lead to a
corresponding change of $\boldsymbol L$ and ${\boldsymbol \Phi}$.
Therefore, different ordering patterns can be employed to generate
multiple branches for exploiting extra transmit diversity gains.
%Inspired from the Turbo codes, a known permutation of the payload data is carried out inside the interleaver.
Motivated by this, we pre-store the designed transmit patterns both
at the transmitter and the receivers, which means that they are
known permutations. Drawing upon previous design methods in
\cite{Rodrigo} and \cite{Rui}, and considering the nature of
distributed users in MU-MIMO scenarios, the design of transmit
patterns is developed in three steps.

As the total number of users is $K$, we first obtain the different
ordering patterns $\boldsymbol  T_u^{(i)}$ between multiple users by
\begin{align}
\boldsymbol  T_u^{(1)}=\boldsymbol I_{K},\\
{\boldsymbol  T_u^{(i)}=\begin {bmatrix} \boldsymbol I_p~~~~ \boldsymbol 0_{p,K-p}\\ \boldsymbol 0_{K-p,p}~~~ \boldsymbol \Pi_{K-p}\\ \end {bmatrix}},~2\leq i\leq K,
\end{align}
where $p=(i-2)$ and $\boldsymbol \Pi_{K-p}$ denotes the exchange
matrix of size $(K-p)\times(K-p)$ with ones on the reverse diagonal
and the superscript $i$ in ${\boldsymbol T_u^{(i)}}$ is termed as
the ordering state. For the $K=3$ case, we have
\begin{align}
{\boldsymbol  T_u^{(1)}=\begin {bmatrix}  1~0~0~ \\  0~1~0~\\0~0~1~ \end {bmatrix}},{\boldsymbol  T_u^{(2)}=\begin {bmatrix}  0~0~1~ \\  0~1~0~\\1~0~0~ \end {bmatrix}},{\boldsymbol  T_u^{(3)}=\begin {bmatrix}  1~0~0~ \\  0~0~1~\\0~1~0~ \end {bmatrix}}.
\end{align}

Next, in order to make the branches as non-contiguous as possible,
we shuffle the streams for each user in a similar way. The ordering
patterns for the $k$th user equipped with $N_{k}$ receive antennas
is given by
\begin{align}
\boldsymbol  T_{s_k}^{(1)}=\boldsymbol I_{N_{k}},\\
{\boldsymbol T_{s_k}^{(j)}=\begin {bmatrix} \boldsymbol I_q~~~~ \boldsymbol 0_{q,N_{k}-q}\\ \boldsymbol 0_{N_{k}-q,q}~~~ \boldsymbol \Pi_{N_{k}-q}\\ \end {bmatrix}},~2\leq j\leq J,
\end{align}
where $q=(j-2)$ and $J$ is the maximum number of ordering states.
Assuming that the first, second, and third user are equipped with
$2$, $2$, and $3$ receive antennas, respectively, then, we have
\begin{align}
{\boldsymbol T_{s_1}^{(1)}={\boldsymbol T_{s_2}^{(1)}}=\begin {bmatrix}  1~0~ \\ 0~1~ \\ \end {bmatrix}},{\boldsymbol T_{s_1}^{(2)}={\boldsymbol T_{s_2}^{(2)}}=\begin {bmatrix}  0~1~ \\ 1~0~ \\ \end {bmatrix}}, \nonumber \\
{\boldsymbol  T_{s_3}^{(1)}=\begin {bmatrix}  1~0~0~ \\  0~1~0~\\0~0~1~ \end {bmatrix}},{\boldsymbol  T_{s_3}^{(2)}=\begin {bmatrix}  0~0~1~ \\  0~1~0~\\1~0~0~ \end {bmatrix}},{\boldsymbol  T_{s_3}^{(3)}=\begin {bmatrix}  1~0~0~ \\  0~0~1~\\0~1~0~ \end {bmatrix}}.
\end{align}
Unlike the ordering states in ${\boldsymbol T_u^{(i)}}$, the total
number of ordering states in ${\boldsymbol T_{s_k}^{(j)}}$ for each
user is not uniform. We first select the user with the maximum
number of receive antenna, which is equal to the maximum ordering
states, i.e., $J={\rm Max}_{k} (N_{k})$ but we note that different
strategies for choosing $J$ are possible.

Finally, we need to package the two ordering patterns $\boldsymbol T_u^{(i)}$ and ${\boldsymbol T_{s_k}^{(j)}}$ together to generate the resulting transmit pattern $\boldsymbol T^{(l)}$. The packaging scheme is that for ordering pattern ${\boldsymbol T_u^{(i)}}$, \revn{the ordering state $j$ is incremented by one while $j\leq J$}.
Inside each $(\boldsymbol T_u^{(i)},{\boldsymbol T_{s_k}^{(j)}})$th packaging process, in order to put ${\boldsymbol T_{s_k}^{(j)}}$ in the right position, we locate the row indices of the nonzero entries in the sparse matrix $\boldsymbol T_u^{(i)}$. Then, we put the ordering pattern ${\boldsymbol T_{s_k}^{(j)}}$ to its corresponding nonzero element in the sparse matrix $\boldsymbol T_u^{(i)}$ and preserve the original sparse pattern. Taking the combination of $(\boldsymbol T_u^{(2)}, {\boldsymbol T_{s_k}^{(2)}})$ for example, the resulting transmit pattern is
\begin{align}
\boldsymbol T^{(2)}=\left [\begin {array}{c c c} \boldsymbol 0&\boldsymbol 0&{{\boldsymbol T_{s_3}^{(2)}}}\\ \boldsymbol 0&{ {\boldsymbol T_{s_2}^{(2)}}}&\boldsymbol 0\\{{\boldsymbol T_{s_1}^{(2)}}}&\boldsymbol 0& \boldsymbol 0\end {array} \right].
\end{align}
For the users equipped with the same number of receive antennas, the total number of ordering states for each user is the same and ${\boldsymbol  T_{s_1}^{(j)}}={\boldsymbol  T_{s_2}^{(j)}}={\boldsymbol  T_{s_3}^{(j)}}$. Then, we use ${\boldsymbol  T_s^{(j)}}$ to denote the ordering patterns for the users and the packaging strategy is simplified by directly implementing the Kronecker product between ${\boldsymbol T_u^{(i)}}$ and ${\boldsymbol  T_{s}^{(j)}}$
\begin{align}
\boldsymbol T^{(l)}={\boldsymbol  T_u^{(i)}}\otimes {\boldsymbol T_{s}^{(j)}},~1\leq l\leq L_B.
\end{align}
With the transmit patterns, a list of transmission branches is constructed.
Then, a proper selection criterion is developed below to find the branch with the minimum sum of errors among all the branches.
The corresponding equivalent channel matrix for a chosen transmit pattern is
denoted as $\boldsymbol H^{(o)}=\boldsymbol T^{(o)}\boldsymbol H$.
Since we employ the MB strategy to
generate extra branches for selection, the BER performance
of the proposed MB-THP algorithms will stay the same
or have a better performance than the conventional THP algorithms.

The maximum number of branches $L_B$ can be equal to $K!J!$,
however, we restrict the total number of branches to no more than
$K\cdot J$ by setting $J={\rm Max}_{k} (N_{k})$. Thus, a reasonable system complexity is maintained.
It is also not necessary to set $L_B$ equal to the maximum number of branches. MB-cTHP and MB-dTHP can approach the performance with $L_B$ branches by using only $2$ or $4$ branches as will be illustrated in Section V. \revns{A total of $L_B$ branches \revnss{is} stored at both the transmitter and the receivers, which requires extra memory for storage. A search procedure is also required to select the best pattern for each transmission.}

\subsection{Selection Criterion for the MB-THP}

From the analysis following equations (14) and (15), the multiplication of different transmit patterns $\boldsymbol T^{(l)}$ by the row vectors of the channel matrix $\boldsymbol H$ results in different error covariance matrices for MB-cTHP and MB-dTHP. For each layer of MB-dTHP, its SNR is inversely proportional to $1/l_{i,i}^2$. For MB-cTHP, it is inversely proportional to $\sum_{i=1}^{S}(1/l_{i,i}^2)$.
Thus, a minimum error selection criterion (MESC) is developed for both MB-cTHP and MB-dTHP to select the best branch according to
\begin{align}
l^{(o)}={\rm arg} \min _{1\leq l\leq L_B} \sum_{1\leq i\leq S}(1/{l}_{i,i}^{(l)})^2,
\end{align}
where $l^{(o)}$ is the selected branch. Then, the received signal ${\boldsymbol r}^{(o)}$ is obtained by
\begin{eqnarray}
{ {\boldsymbol r}}^{{(o)}^{\rm (dTHP)}}&=&\boldsymbol G^{(o)}({\boldsymbol H^{(o)}}\boldsymbol F^{(o)}{\boldsymbol x^{(o)}} +\boldsymbol n),\\
{ {\boldsymbol r}}^{{(o)}^{\rm (cTHP)}}&=&\beta({\boldsymbol H^{(o)}}\cdot{1\over\beta}\boldsymbol F^{(o)}\boldsymbol G^{(o)}{\boldsymbol x^{(o)}} +\boldsymbol n).
\end{eqnarray}
Since the transmit patterns are pre-stored and known both at the transmit and
receive terminals, the transmitter can inform the receiver about the
index of the selected pattern or the receiver can search for the
best pattern. Then, the ordered signal ${\boldsymbol r}^{(o)}$ is transformed back to ${\boldsymbol r}$ by
$\boldsymbol T^{{(o)}^T}$ at each receive terminal. Next, the
transformed signal ${\boldsymbol r}$ is passed through the
modulo processing to remove the offset by the perturbation vector
$\boldsymbol d^{(o)}$, and a quantization function is followed to
slice the symbols to the nearest points of the constellation as
\begin{align}
 \hat {\boldsymbol s}={\rm Q}({\rm M}({\boldsymbol r})),
\end{align}
where ${\rm Q}(\cdot)$ is the slicing function and ${\rm M}(\cdot)$
is the modulo operation implemented element-wise as in (8).

\subsection{Derivation of Filters for the MB-MMSE-THP}
It is well-known that MMSE based precoding algorithms always have a better performance than that of ZF based.
The filters of the cTHP based MMSE design are deduced from an optimization problem in \cite{Joham01, Joham02}, which results in a high computational complexity since multiple calculations of matrix inverses are required.
The orthogonality principle is utilized in \cite{Jia01} to obtain the filters of  MMSE-dTHP.
In \cite{Habendorf}, the filters of MMSE-cTHP are derived from an extended system model, which
is simpler and more effective compared to the above two methods because the LQ decomposition is utilized. The receive model for MMSE-cTHP based on the extended matrix, however, is not given in \cite{Habendorf}. In this work, we derive the filters of the proposed MB-MMSE-cTHP and MB-MMSE-dTHP based on the extended matrix and their corresponding receive models are also described.

Define the $N_r\times(N_r+N_t)$ extended channel matrix $\boldsymbol {\underline H}$ for the MB-MMSE
precoding schemes as
\begin{align}
{\underline {\boldsymbol H}^{(l)}}=\left[\begin {array}{c c} {\boldsymbol H^{(l)}},&\sigma_n \boldsymbol I_{N_r}\end {array}\right],
\end{align}
where ${\boldsymbol H^{(l)}}=\boldsymbol T^{(l)}\boldsymbol H$.
Then, the linear precoding MMSE filter can be rewritten as $\boldsymbol P_{\rm MMSE}^{(l)}=\boldsymbol A {{\underline {\boldsymbol H}^{(l)}}}^H({\underline {\boldsymbol H}^{(l)}}{{\underline {\boldsymbol H}^{(l)}}}^H)^{-1}$, where $\boldsymbol A=[\boldsymbol I_{N_t},\  \boldsymbol 0_{N_t,N_r}]$. By
implementing the LQ decomposition of the extended channel matrix
${\underline {\boldsymbol H}^{(l)}}$ we have
\begin{align}
{\underline {\boldsymbol H}^{(l)}}={\underline {\boldsymbol L}^{(l)}} {\underline {\boldsymbol Q}^{(l)}}
={\underline {\boldsymbol L}^{(l)}}\left[\begin {array}{c c} {{\boldsymbol Q^{(l)}_1}},&{\boldsymbol Q^{(l)}_2}\end {array}\right],
\end{align}
where ${\underline {\boldsymbol L}^{(l)}}$ is a $N_r\times N_r$ lower triangular matrix and the $N_r\times (N_r+N_t)$ matrix ${\underline {\boldsymbol Q}^{(l)}}$ with orthogonal columns can be partitioned into the $N_r\times N_t$ matrix ${\boldsymbol Q^{(l)}_1}$ and the $N_r\times N_r$ matrix ${\boldsymbol Q^{(l)}_2}$. From (29) and (30), the following relations hold
\begin{align}
{\boldsymbol H^{(l)}}&={\underline {\boldsymbol L}^{(l)}} {\boldsymbol Q^{(l)}_1},\\
{\underline {\boldsymbol L}^{(l)}}^{-1}&={1\over \sigma_n} {\boldsymbol Q^{(l)}_2},\\
\boldsymbol A{\underline {\boldsymbol Q}^{(l)}}^H&={\boldsymbol
Q_1^{(l)}}^H.
\end{align}
Therefore, the filters for the MB-MMSE-cTHP and the MB-MMSE-dTHP schemes can be obtained as
\begin{align}
\boldsymbol F^{(l)}&= {\underline {\boldsymbol Q}^{(l)}}^H,\\
\boldsymbol G^{(l)}&={\rm {diag}}[\underline l_{1,1}^{(l)},\underline l_{2,2}^{(l)},\cdots,\underline l_{N_t,N_t}^{(l)}]^{-1},\\
\boldsymbol B^{{(l)}^{(\rm dTHP)}}&=\boldsymbol G^{(l)} {\underline {\boldsymbol L}^{(l)}},\\
\boldsymbol B^{{(l)}^{(\rm cTHP)}}&={\underline {\boldsymbol L}^{(l)}}\boldsymbol G^{(l)},
\end{align}
where $\underline l_{ii}^{(l)}$ are the diagonal elements of
${\underline {\boldsymbol L}^{(l)}}$. The received signal for the
$l$th branch is
\begin{align}
 {\boldsymbol r^{{(l)}^{(\rm dTHP)}}}&=\boldsymbol G^{(l)}({\boldsymbol H^{(l)}}\boldsymbol A\boldsymbol F^{(l)}{\boldsymbol x^{(l)}} +\boldsymbol n), \\
 {\boldsymbol r^{{(l)}^{(\rm cTHP)}}}&=\beta({\boldsymbol H^{(l)}}\cdot{1\over\beta}\boldsymbol A\boldsymbol F^{(l)}\boldsymbol G^{(l)}{\boldsymbol x^{(l)}}+\boldsymbol n).
\end{align}
It is worth noting that the multiplication by $\boldsymbol A$ will not result in transmit power amplification since $\boldsymbol A\boldsymbol A^H=\boldsymbol I_{N_t}$ ($\boldsymbol A$ is pseudo-unitary).
The implementation steps of the MB-MMSE-THP algorithms are summarized in Table I.
%As MB-MMSE-THP does not only take the interference but also the noise into account, a better BER performance can be achieved compared to the MB-ZF-THP algorithm. Therefore, we focus on the MB-MMSE-THP structures in this work.
\begin{table}[htp]
\caption{Proposed MB-MMSE-THP Algorithms} % title of Table
\centering % used for centering table
\begin{tabular}{l l} % centered columns (4 columns)
\hline\hline %inserts double horizontal lines
Steps & Operations \\ [0.5ex] % inserts table
%heading
\hline % inserts single horizontal line
& {\bf Compute the extended channel matrix for the $l$th branch}\\
(1)& ${\underline {\boldsymbol H}^{(l)}}=\left[\begin {array}{c c} {\boldsymbol T^{(l)}} {\boldsymbol H},&\sigma_n \boldsymbol I_S\end {array}\right]$\\
& {\bf Implement the LQ decomposition}~~~~~~~~~~~~~~~\\
(2)& ${\underline {\boldsymbol H}^{(l)}}={\underline {\boldsymbol L}^{(l)}} {\underline {\boldsymbol Q}^{(l)}}$\\
& {\bf Obtain the filters for MB-cTHP and MB-dTHP}~~~~~~~~~~~~~~~\\
(3)& $\boldsymbol F^{(l)}= {\underline {\boldsymbol Q}^{(l)}}^H, \boldsymbol G^{(l)}={\rm {diag}}[\underline l_{1,1}^{(l)},\underline l_{2,2}^{(l)},\cdots,\underline l_{N_t,N_t}^{(l)}]^{-1}$,\\
& $ \boldsymbol B^{{(l)}^{(\rm cTHP)}}={\underline {\boldsymbol L}^{(l)}}\boldsymbol G^{(l)}, ~\boldsymbol B^{{(l)}^{(\rm dTHP)}}=\boldsymbol G^{(l)} {\underline {\boldsymbol L}^{(l)}}$\\
& {\bf The MESC selection criterion}~~~~~~~~~~~~~~~\\
(4)& for j~=~1~:~${L_B}^\ddag$ ~\ddag ~${L_B}$ is the total number of branches\\
(5)&~~~${\rm MESC}(j) =\sum_{i=1}^{S}(1/l_{i,i}^2)$\\
(6)&end\\
(7)& ${l^{(o)}}^{\ddag\ddag}={\rm Min} ({\rm MESC})$ ~\ddag\ddag ~${l^{(o)}}$ is the selected optimal branch\\
& {\bf The successive cancelation process }~~~~~~~~~~~~~~~\\
(8)& for i~=~1~:~$S$\\
(9)&~~~$x^{(o)}(i) =s_i-\sum _{j\neq i}^{S} b_{i,j} x^{(o)}(j)$\\
(10)&~~~$x^{(o)}(i)  = {\rm M}(x^{(o)}(i))$\\
(11)&end\\
& {\bf The received signal}\\
(12)& $\beta={{\rm E}\| {\boldsymbol F\boldsymbol G\boldsymbol x}\|\over{\rm E}\| {\boldsymbol s}\|}$\\
(13)& ${\boldsymbol r}^{{(o)}^{\rm (cTHP)}}=\beta({\boldsymbol H^{(o)}}\cdot{1\over\beta}\boldsymbol F^{(o)}\boldsymbol G^{(o)}{\boldsymbol x^{(o)}} +\boldsymbol n)$\\
(14)& ${\boldsymbol r}^{{(o)}^{\rm (dTHP)}}=\boldsymbol G^{(o)}({\boldsymbol H^{(o)}}\boldsymbol F^{(o)}{\boldsymbol x^{(o)}} +\boldsymbol n)$\\
(15)& ${\hat {\boldsymbol s}}^{\rm (cTHP)}={\rm Q}({\rm M}(\boldsymbol T^{{(o)}^T}{\boldsymbol r}^{{(o)}^{\rm (cTHP)}}))$\\
(16)& ${\hat {\boldsymbol s}}^{\rm (dTHP)}={\rm Q}({\rm M}(\boldsymbol T^{{(o)}^T}{\boldsymbol r}^{{(o)}^{\rm (dTHP)}}))$\\
% &{\color{red}\ddag ~${L_B}$ is the total number of branches}\\
% &{\color{red}\ddag ~${l^{(o)}}$ is the selected optimal branch}\\
[1ex] % [1ex] adds vertical space
\hline %inserts single line
\end{tabular}
\end{table}

\section{Performance Analysis}
In this section, we consider a performance analysis in terms of error covariance, sum-rate, and computational complexity.
\subsection{Performance Analysis of the Error Covariance Matrix}
 The autocorrelation matrices of the interference-plus-noise power in ZF-dTHP and ZF-cTHP have been given in \cite{MH}, however, the comparison has not been done. In this section, we illustrate the BER performances in terms of error covariance.
For the comparison between ZF-dTHP and ZF-cTHP, we assume $i$ is an arbitrary layer, then from equations (14) and (15) we have
\begin{align}
{{\boldsymbol \Phi}^{l^{(o)}}_{{\rm ZF-cTHP}_{i,i}}\over {\boldsymbol \Phi}^{l^{(o)}}_{{\rm ZF-dTHP}_{i,i}}}=1+{l^{l^{(o)}}_{i,i}}^2\sum_{j\neq i}^{S}(1/{l^{l^{(o)}}_{j,j}}^2).
\end{align}
That is, $\forall~i:~{{\boldsymbol \Phi}^{l^{(o)}}_{{\rm ZF-dTHP}_{i,i}}<{\boldsymbol \Phi}^{l^{(o)}}_{{\rm ZF-cTHP}_{i,i}}}$. Since the BER performance is largely related to the error covariance matrix, we expect a better BER performance achieved by ZF-dTHP over ZF-cTHP. This is also verified by the simulation result in \cite{MH}, from which a slightly better BER performance of ZF-dTHP over ZF-cTHP is reported.

The comparison between MMSE-dTHP and MMSE-cTHP, however, has not been analyzed nor simulated in the literature so far. Substituting (31), (33), (34) and (36) into (38), we can get the error covariance matrix for MMSE-dTHP as
\begin{align}
{\boldsymbol \Phi}^{(l)}_{\rm {MMSE-dTHP}}&=&{\rm diag}(\sigma_n/{\underline l^{(l)}_{1,1}},\cdots,\sigma_n/{\underline l^{(l)}_{S,S}})^2.
\end{align}
For the MMSE-cTHP we start from the calculation of $\beta$ for a more accurate expression by
\begin{align}
\beta^2={{\rm E}\| {\boldsymbol A\boldsymbol F^{(l)}\boldsymbol G^{(l)}\boldsymbol x^{(l)}}\|^2\over\sigma_s^2},
\end{align}
where $\sigma_s^2={\rm E}\| {\boldsymbol s}\|^2$.
Since $\boldsymbol x^{(l)}=\boldsymbol {B^{(l)}}^{-1}\boldsymbol v^{(l)}$, $\boldsymbol B^{(l)}={\underline {\boldsymbol L}^{(l)}}\boldsymbol G^{(l)}$ and ${\underline {\boldsymbol L}^{(l)}}^{-1}={1\over \sigma_n} {\boldsymbol Q^{(l)}_2}$, the multiplication $\boldsymbol A\boldsymbol F^{(l)}\boldsymbol G^{(l)}\boldsymbol x^{(l)}$ is obtained as
\begin{align}
\boldsymbol A\boldsymbol F^{(l)}\boldsymbol G^{(l)}\boldsymbol x^{(l)}={1\over \sigma_n}\boldsymbol A\boldsymbol F^{(l)}{\boldsymbol Q^{(l)}_2\boldsymbol v^{(l)}}.
\end{align}
Then, by applying the equivalence ${\rm tr}(\boldsymbol A\boldsymbol B\boldsymbol C)={\rm tr}(\boldsymbol C\boldsymbol A\boldsymbol B)$, the normalization factor $\beta$ can be expressed as
\begin{align}
\beta^2={{\sigma_v^{(l)}}^2\over\sigma_n^2\sigma_s^2},
\end{align}
where the quantity ${\sigma_v^{(l)}}^2$ is the variance of $\boldsymbol v^{(l)}$. Therefore, the error covariance matrix for MMSE-cTHP is obtained as
\begin{align}
{\boldsymbol \Phi}^{(l)}_{\rm {MMSE-cTHP}}={\rm diag}\biggl({{\sigma_v^{(l)}}^2\over\sigma_s^2},\cdots,{{\sigma_v^{(l)}}^2\over\sigma_s^2}\biggr).
\end{align}
By changing the transmit signal order, different perturbation vectors $\boldsymbol d^{(l)}$ are obtained in MB-MMSE-cTHP.
The multi-branch processing is actually used to select the one with the minimum ${\sigma_v^{(l)}}^2$ among all the $L_B$ branches in MB-MMSE-cTHP algorithms.

For the comparison between MB-MMSE-dTHP and MMSE-dTHP, we have the proposition below.

\revn{{\it Proposition 1:}} The trace of the error covariance matrix for the proposed MB-MMSE-dTHP technique is upper bounded by that of the conventional MMSE-dTHP scheme, i.e.,
\begin{align}
{\rm tr}({{\boldsymbol \Phi}_{\rm {MB-MMSE-dTHP}})\leq {\rm tr}({\boldsymbol \Phi}_{\rm {MMSE-dTHP}}}).
% {{\boldsymbol \Phi}^{l^{(o)}}_{\rm {MB-MMSE-dTHP}}\leq{\boldsymbol \Phi}_{\rm {MMSE-dTHP}}}
\end{align}

{\it Proof:} From the MESC selection criterion in (25), the selected branch $l^{(o)}$ corresponds to the sum of the elements associated with the smallest value, i.e.,
\begin{align}
{\rm tr}({\boldsymbol \Phi}_{\rm {MB-MMSE-dTHP}})=\sum_{1\leq i\leq S}(1/\underline {l}_{i,i}^{(o)})^2.
\end{align}
With the MESC selection criterion, we have
\begin{align}
\sum_{1\leq i\leq S}(1/\underline {l}_{i,i}^{(o)})^2\leq\sum_{1\leq i\leq S}(1/\underline {l}_{i,i}^{(l)})^2, l=1,2,\cdots,L_B.
\end{align}
By writing the above quantities without the sum, we get
\begin{align}
\biggl({1\over{\underline {l}_{1,1}^{(o)}}}\biggr)^2+\cdots+\biggl({1\over{\underline {l}_{S,S}^{(o)}}}\biggr)^2\leq
\biggl({1\over{\underline {l}_{1,1}^{(l)}}}\biggr)^2+\cdots+\biggl({1\over{\underline {l}_{S,S}^{(l)}}}\biggr)^2, \nonumber\\
\biggl[\biggl({1\over{\underline {l}_{1,1}^{(o)}}}\biggr)^2-\biggl({1\over{\underline {l}_{1,1}^{(l)}}}\biggr)^2\biggr]+\cdots+\biggl[\biggl({1\over{\underline {l}_{S,S}^{(o)}}}\biggr)^2-\biggl({1\over{\underline {l}_{S,S}^{(l)}}}\biggr)^2\biggr]\leq0.
\end{align}
If we choose $\underline {l}_{i,i}^{(o)}$ to be identical to $\underline {l}_{i,i}^{(l)}$ then we prove the equality ${\rm tr}({{\boldsymbol \Phi}_{\rm {MB-MMSE-dTHP}})={\rm tr}({\boldsymbol \Phi}_{\rm {MMSE-dTHP}}})$.
If we choose at least one element $\underline {l}_{i,i}^{(o)}>\underline {l}_{i,i}^{(l)}~ {\rm or}~\underline {l}_{i,i}^{(o)}-\underline {l}_{i,i}^{(l)}=\epsilon$ while keeping the others identical $\underline {l}_{j,j}^{(o)}=\underline {l}_{j,j}^{(l)}, ~j\neq i$ then we prove the inequality
${\rm tr}({{\boldsymbol \Phi}_{\rm {MB-MMSE-dTHP}})<{\rm tr}({\boldsymbol \Phi}_{\rm {MMSE-dTHP}}})$, where $\epsilon$ is a small real positive value \revn{$\hfill\square$}.

For MMSE-cTHP, the overall SNR performance is influenced by the sum of each layer, then from (45) we have
\begin{align}
{\rm tr}({\boldsymbol \Phi}_{\rm {MMSE-cTHP}})={K{\sigma_v^{(l)}}^2\over\sigma_s^2},\\\nonumber
{\rm tr}({\boldsymbol \Phi}_{\rm {MB-MMSE-cTHP}})={K{\sigma_v^{(o)}}^2\over\sigma_s^2}.
\end{align}
Because of the MESC selection process, we have obtained that $\forall~l: {\sigma_v^{(o)}}^2\leq{\sigma_v^{(l)}}^2$.
Thus, it is straightforward to conclude that
\begin{align}
{\rm tr}({\boldsymbol \Phi}_{\rm {MB-MMSE-cTHP}})\leq{\rm tr}({\boldsymbol \Phi}_{\rm {MMSE-cTHP}}).
\end{align}

Therefore, we expect that a better BER performance can be achieved by the proposed MB-dTHP and MB-cTHP, respectively, as compared to their original counterparts.

\subsection{Sum-Rate Performance Analysis}
From the analysis illustrated in Section III, the MU-MIMO channel is decomposed into parallel AWGN channels in the THP systems. Therefore, the $i$th SNR for the $l$th branch transmit signal of MB-ZF-THP is given by \revn{\cite{Tse, MH}}
\begin{eqnarray}
{{\gamma_i}^{(l)}}^{\rm (MB-ZF-dTHP)}&=&{\sigma_s^2\over\sigma_n^2(1/ {l^{(l)}_{i,i}}^2)},\\
{{\gamma_i}^{(l)}}^{\rm (MB-ZF-cTHP)}&=&{\sigma_s^2\over\sigma_n^2 \sum_{i=1}^{S}(1/{l^{(l)}_{i,i}}^2)}.
\end{eqnarray}
Then, the achievable sum rates for the $l$th branch of MB-ZF-dTHP and MB-ZF-cTHP are respectively given by
\begin{eqnarray}
C^{(l)}_{\rm (MB-ZF-dTHP)}&=&\sum_{i=1}^S{\rm log}\Bigl(1+{\sigma_s^2{l^{(l)}_{i,i}}^2\over\sigma_n^2}\Bigr),\\
C^{(l)}_{\rm (MB-ZF-cTHP)}&=&S~{\rm log}\Biggl(1+{\sigma_s^2\over\sigma_n^2 \sum_{i=1}^{S}(1/{l^{(l)}_{i,i}}^2)}\Biggr).
\end{eqnarray}
\revn{From \revns{equations} (41) and (42), the achievable sum rates of MB-MMSE-dTHP and MB-MMSE-cTHP can be expressed, \revns{respectively}, as follows}
\begin{eqnarray}
C^{(l)}_{\rm (MB-MMSE-dTHP)}&=&\sum_{i=1}^S{\rm log}\Bigl(1+{\sigma_s^2\underline {l^{(l)}_{i,i}}^2\over\sigma_n^2}\Bigr),\\
C^{(l)}_{\rm (MB-MMSE-cTHP)}&=&S~{\rm log}\Biggl(1+{\sigma_s^4\over{\sigma_v^{(l)}}^2}\Biggr).
\end{eqnarray}

%By constructing a diagonal matrix $\boldsymbol V^{(l)}\in\mathbb{C}^{S\times S}$ with $\sigma_s {l^{(l)}_{i,i}}\over\sigma_n$ on the $i$th diagonal entry, we can rewrite the sum rates of dTHP as
%\begin{eqnarray}
%C^{(l)}_{\rm (ZF-dTHP)}&=&{\rm log}\Bigl({\rm det}(\boldsymbol I+{\boldsymbol V}^{(l)}{{\boldsymbol V}^{(l)}}^H)\Bigr),\\
%C^{(l)}_{\rm (MMSE-dTHP)}&=&{\rm log}\Bigl({\rm det}(\boldsymbol I+{\underline{\boldsymbol V}}^{(l)}{{\underline{\boldsymbol V}}^{(l)}}^H)\Bigr).
%\end{eqnarray}
%From the matrix theory \cite{MatrixTheo}, for $l=1,2,\cdots,L_B$ we have
%\begin{eqnarray}
%{\rm det}({\underline{\boldsymbol V}}^{(1)}{{\underline{\boldsymbol V}}^{(1)}}^H)=\cdots={\rm det}({\underline{\boldsymbol V}}^{(L_B)}{{\underline{\boldsymbol V}}^{(L_B)}}^H).
%\end{eqnarray}
From (54) and (56), the difference of the overall average SNR for the $l$th branch is small. Thus, we expect that MB-MMSE-dTHP with different branches shares a similar sum-rate performance. For MB-MMSE-cTHP, the ${\sigma_v^{(o)}}^2$ of the selected $l^{(o)}$th branch has the minimum value among all the branches because of the multi-branch processing and the selection, that is
%  For MB-MMSE-cTHP, however, the variance of the effective transmit signal ${\sigma_v^{(l)}}^2$ is dependent on the ordering.
\begin{align}
{\sigma_v^{(o)}}^2\leq{\sigma_v^{(l)}}^2, ~l=1,\cdots,L_B,
\end{align}
Thus, we have
\begin{align}
C_{\rm (MMSE-cTHP)}\leq C^{(o)}_{\rm (MB-MMSE-cTHP)},
\end{align}
which means the sum-rate performance of MMSE-cTHP can be improved by the proposed MB-MMSE-cTHP algorithm.

\subsection{Complexity Analysis}
In this section we use the total number of floating point operations (FLOPs) to measure the
computational complexity of the proposed and existing algorithms.
The number of FLOPs for the LQ decomposition is obtained by assuming that the LQ decomposition is computed by using the Householder transformation given in \cite{MatrixTheo}.
We summarize the total number of FLOPs needed for the matrix
operations below:
\begin {itemize}
\item {Multiplication of $m\times n$ and $n\times p$ complex matrices: $8mnp-2mp$;}

\item {LQ decomposition of an $m\times n~(m\leq n)$ complex matrix: $8m^2(n-{1\over 3}m)$;}

\item {Pseudo-inversion of an $m\times n$ complex matrix: $({4\over 3}m^3+7m^2n-m^2-2mn)$}.
\end {itemize}

The number of FLOPs needed for BD and RBD can be found in \cite{Zu,Zu02}. The computational complexity of MMSE-THP based on multiple matrix inversions in \cite{Joham01} has been given in \cite{Kusume}.
The complexity reported in \cite{Kusume}, however, is only computed in terms of the number of multiplications and additions.
For the complex multiplications and additions, it respectively needs $6$ and $2$ FLOPs.
Thus, the number of FLOPs needed by MMSE-THP in \cite{Joham01} is at least $24N_r^4+48N_r^3+N_tN_r$. For MMSE-THP based on the Cholesky factorization in \cite{Kusume}, the number of FLOPs needed is at least ${20\over 3}N_r^3+8N_r^2N_t$.
The ZF-VP in \cite{Hochwald} and MMSE-VP in \cite{MMSE-VP} are implemented by using the sphere decoder (SD) algorithm which is employed for sphere encoding.
The complexity of SD is associated with the constellation size $M$ and the radius $d$ which is chosen to be a scaled version of the noise variance \cite{SD}.
The required multiplications and additions of SD are given in \cite{Keke02}.

For simplicity, we assume that the number of transmit antennas
$N_t$ and the number of receive antennas $N_r$ are equal to $n$.
From the above derivation, MB-MMSE-dTHP and MB-MMSE-cTHP share the same computational complexity.
The number of FLOPs for the above precoding algorithms are listed in Table II, where $\bar N_k=n-N_k$.
In case of system dimension $n=6$, number of users $K=3$, each user equipped with $N_k=2$ receive antennas and number of branches $L_B=2$,
the required number of FLOPs of MB-ZF-THP and MB-MMSE-THP is much lower than the BD, RBD, conventional MMSE-THP in \cite{Joham01} and VP algorithms.

\begin{table}[!t]
\caption{Comparison of the complexity} % title of Table
\centering % used for centering table
\begin{tabular}{l l r} % centered columns (4 columns)
\hline\hline %inserts double horizontal lines
Algorithm &FLOPs &Case\\
% & &$L_B=2$\\ [0.5ex] % inserts table
%heading
\hline % inserts single horizontal line
ZF & $16n^3+3n^2-2n$ & $3552$\\

MMSE & $16n^3+3n^2$ & $3564$\\

BD & $K(72N_k^3+72n_i^2n+32N_kn^2$ &\\
 & $-2N_k^2+32n{\bar N_k}^2+64{\bar N_k}^3)$ & $35304$\\

RBD & $K(72N_k^3+72N_k^2n+32N_kn^2-2N_k^2$ &\\
 & $+8n^3+18n+\bar n_i+32n{\bar N_k}^2+64{\bar N_k}^3)$ & $40824$\\

ZF-THP &${40\over3}n^3+10n^2+22n$& $3372$\\

MMSE-THP \cite{Joham01}& $24n^4+48n^3+n^2$ &41508\\

MMSE-THP &${64\over3}n^3+10n^2+22n$& $5100$\\

MB-ZF-THP &$L_B({40\over3}n^3+10n^2+22n)$& $6744$\\

MB-MMSE-THP & $L_B({64\over3}n^3+10n^2+22n)$ &$10200$\\

ZF-VP/MMSE-VP& $8\sum_{k=1}^n{Mk\pi^{k\over2}\over \Gamma(k/2+1)}d^k$ &\\
& $~+16n^2-2n+4$ & $4.8\cdot10^{7}$
% Low-MB-MMSE-THP & $L_c({64\over3}n^3+10n^2+22n)$ &$5100\times L_c$
\\[1ex] % [1ex] adds vertical space
\hline %inserts single line
\end{tabular}
\end{table}

The required number of FLOPs of the proposed and existing algorithms is simulated for different system dimensions and the results are depicted in Fig. 3. It is clear that VP shows the highest complexity. The computational cost of BD, RBD, and MMSE-THP in \cite{Joham01} is relatively high compared to the proposed MB-MMSE-THP algorithms due to multiple SVD or matrix inversion operations are implemented. Moreover, the proposed MB-MMSE-THP algorithm with $L_B=2$ and $L_B=4$ branches has a complexity that is slightly higher than the ZF-THP, MMSE, and MMSE-THP algorithms especially when the system dimension is below $10$.
\begin{figure}[htp]
\centering
\def\epsfsize#1#2{1.0\columnwidth}
\epsfbox{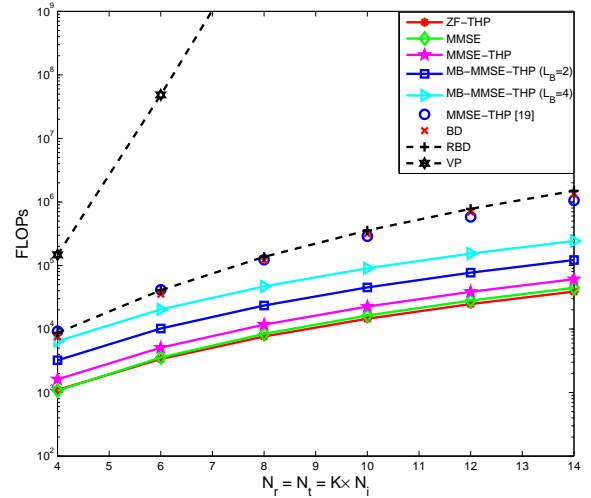} \vspace{-1.25em} \caption{\footnotesize
Complexity Analysis (The proposed MB-cTHP and MB-dTHP share the same
complexity).}
\end{figure}

\section{Simulation Results}
In this section, we assess the performance of the proposed MB-THP algorithms. A system with $N_t=8$ transmit antennas and $K=4$
users each equipped with $N_k=2$ receive antennas is considered;
this scenario is denoted as the $(2,2,2,2)\times 8$ case. The quantity $E_b/N_0$ is defined as $E_b/N_0={N_r E_s\over N_tN\sigma_n^2}$ with
$N$ being the number of information bits transmitted per channel
symbol. Uncoded QPSK and 16-QAM modulation schemes are employed in the simulations.
The channel matrix ${\boldsymbol H}$ is assumed to be a
complex i.i.d. Gaussian matrix with zero mean and unit variance.
% Perfect CSI is considered first, and then the influence of imperfect CSI is evaluated.
The number of branches employed for MB-MMSE-THP is $L_B=2,4,6,8$, respectively.
The number of simulation trials is $10^6$ and the packet length is $10^2$ symbols.
\subsection{Perfect Channel State Information Scenario}
\begin{figure}[!ht]
      \centering
      \subfloat[BER performance of THP, QPSK\label{THP_QPSK}]{
        \includegraphics[scale=0.41]{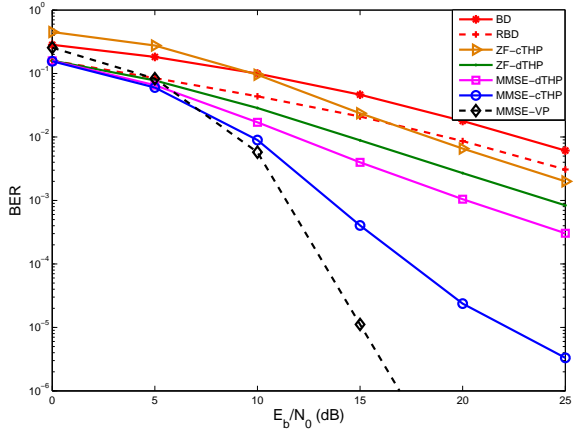}
      }
      \hspace{-10mm}
      \subfloat[BER performance of THP, 16-QAM\label{THP_16QAM}]{
        \includegraphics[scale=0.41]{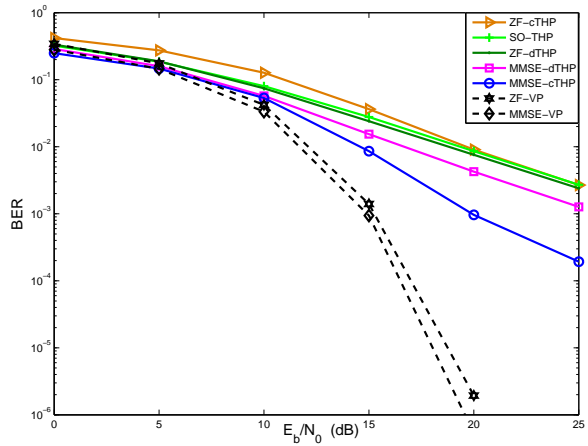}
      }
      \caption{BER performance of THP}\label{THP_BER}
    \end{figure}

As illustrated in Fig. \ref {THP_QPSK}, the BER performance of the BD and RBD precoding algorithms is worse than that of the THP algorithms. For the THP algorithms, a better BER performance is offered by ZF-dTHP over ZF-cTHP since $\forall~i:~{{\boldsymbol \Phi}^{l^{(o)}}_{{\rm ZF-dTHP}_{i,i}}<{\boldsymbol \Phi}^{l^{(o)}}_{{\rm ZF-cTHP}_{i,i}}}$ as we illustrated in (40). However, a much better BER performance is achieved by MMSE-cTHP than MMSE-dTHP, which verifies the analysis developed in Section IV.
%\begin{figure}[htp]
%\begin{center}
%\def\epsfsize#1#2{0.6\columnwidth}
%\epsfbox{All_QPSK_BER.eps} \vspace{-1.25em}
%\caption{\footnotesize BER performance of THP, QPSK.}\label{Fig. 4.}
%\end{center}
%\end{figure}
The comparison among nonlinear precoding algorithms with 16-QAM is displayed in Fig. \ref {THP_16QAM}. The same phenomenon is also observed for the two types of THP with 16-QAM. A slightly better BER performance is offered by ZF-dTHP over ZF-cTHP, whereas, the situation is reversed for MMSE-THP. The THP with successive BD implementation (SO-THP) algorithm in \cite{Veljko2} shows a slightly better performance than ZF-cTHP at low $E_b/N_0$s, however, its performance is almost the same as ZF-dTHP and ZF-cTHP at high $E_b/N_0$s. The maximum transmit diversity order is achieved by ZF-VP and MMSE-VP algorithms.
%\begin{figure}[htp]
%\begin{center}
%\def\epsfsize#1#2{0.6\columnwidth}
%\epsfbox{THP_MU_MIMO_BER_16QAM.eps} \vspace{-1.25em}
%\caption{\footnotesize BER performance of THP, 16-QAM.}\label{Fig. 5.}
%\end{center}
%\end{figure}

\begin{figure}[!ht]
      \centering
      \subfloat[BER performance of cTHP,  $(2,2,2,2)\times 8$ MIMO, 16-QAM.\label{cTHP_16QAM}]{
        \includegraphics[scale=0.41]{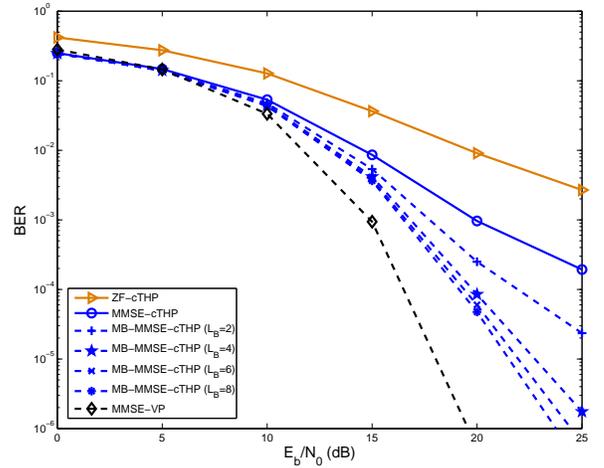}
      }
      \hspace{-10mm}
      \subfloat[BER performance of cTHP,  $(2,2,2,2)\times 8$ MIMO, QPSK.\label{cTHP_QPSK}]{
        \includegraphics[scale=0.41]{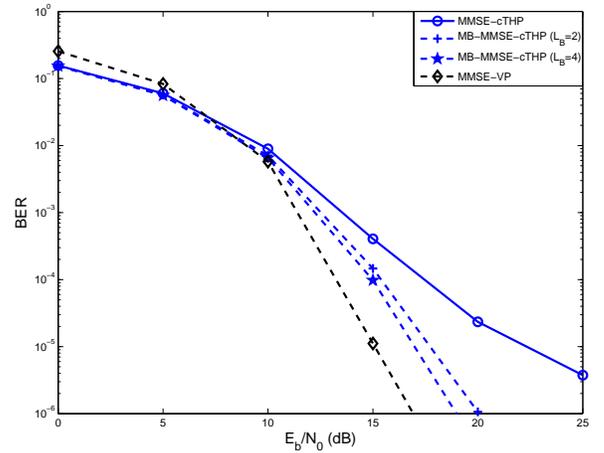}
      }
      \caption{BER performance of cTHP}\label{cTHP_BER}
    \end{figure}

The BER performance of the proposed MB-MMSE-cTHP with 16-QAM and QPSK are shown in Fig. \ref{cTHP_16QAM} and Fig. \ref{cTHP_QPSK}, respectively.
From Fig. \ref{cTHP_16QAM}, the proposed MB-MMSE-cTHP with $L_B=2, 4, 8$ branches has a gain of more than 2 dB, 3 dB, and 3.4 dB as compared
to the conventional MMSE-cTHP and the performance gap between MB-MMSE-cTHP with $L_B=4$ and MMSE-VP is only 2 dB at the BER of $10^{-3}$.
For the QPSK modulation in Fig. \ref{cTHP_QPSK}, the BER performance of MB-MMSE-cTHP with $L_B=4$ is better than MMSE-VP at low $E_b/N_0$s and is very close to that of MMSE-VP at the BER of $10^{-3}$ but requires a much lower computational complexity.
%\begin{figure}[htp]
%\begin{center}
%\def\epsfsize#1#2{0.6\columnwidth}
%\epsfbox{MB_MMSE_cTHP_8X8_16QAM_BER.eps} \vspace{-1.25em}
%\caption{\footnotesize BER performance of cTHP, 16-QAM.}\label{Fig. 6.}
%\end{center}
%\end{figure}

\begin{figure}[htp]
\begin{center}
\def\epsfsize#1#2{1.0\columnwidth}
\epsfbox{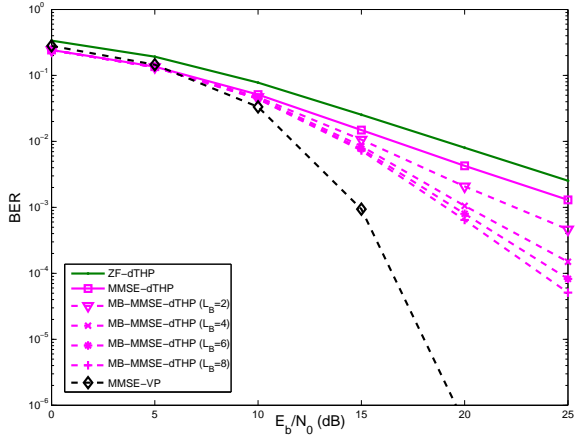} \vspace{-1.25em} \caption{\footnotesize BER
performance of dTHP,  $(2,2,2,2)\times 8$ MIMO,
16-QAM.}\label{dTHP_16QAM}
\end{center}
\end{figure}

Fig. \ref {dTHP_16QAM} displays the BER performance of the proposed
MB-MMSE-dTHP algorithms. For the proposed MB-MMSE-dTHP with $L=2, 4,
8$ branches, there is a gain of more than 3.6 dB, 6 dB, and 7 dB as
compared to the conventional MMSE-dTHP at the BER of $10^{-3}$,
respectively.

As illustrated by Fig. 5 and Fig. 6, the transmit diversity of the proposed MB-MMSE-cTHP and MB-MMSE-dTHP algorithms is between the VP and the conventional MMSE-THP algorithms because a list of branches is constructed and the best candidate is selected by the proposed algorithms. It is worth noting that for both MB-MMSE-cTHP and MB-MMSE-dTHP with only $2$ branches, there is a considerable performance improvement and their BER performances with $4$ branches can approach the one with $8$ branches. Especially for MB-MMSE-cTHP, its BER performance with only $4$ branches is not far from MMSE-VP with much less computational complexity.

%\begin{figure}[htp]
%\begin{center}
%\def\epsfsize#1#2{0.6\columnwidth}
%\epsfbox{MB_MMSE_cTHP_8X8_QPSK_BER.eps} \vspace{-1.25em}
%\caption{\footnotesize BER performance of cTHP, QPSK.}\label{Fig. 7.}
%\end{center}
%\end{figure}
\begin{figure}[!ht]
      \centering
      \subfloat[Sum-rate performance of cTHP, $(2,2,2,2)\times 8$ MIMO, 16-QAM.\label{cTHP_Sumrate}]{
        \includegraphics[scale=0.41]{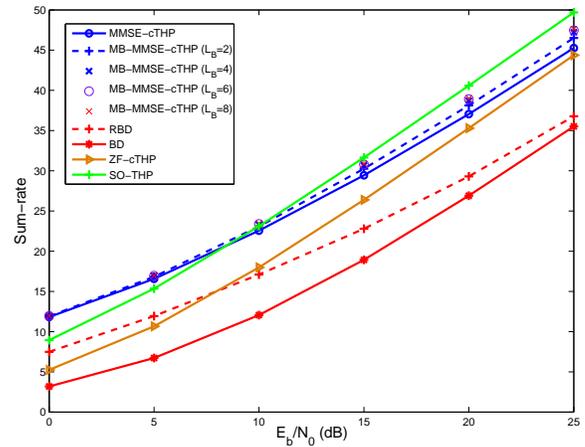}
      }
      \hspace{-10mm}
      \subfloat[Sum-rate performance of dTHP, $(2,2,2,2)\times 8$ MIMO, 16-QAM.\label{dTHP_Sumrate}]{
        \includegraphics[scale=0.41]{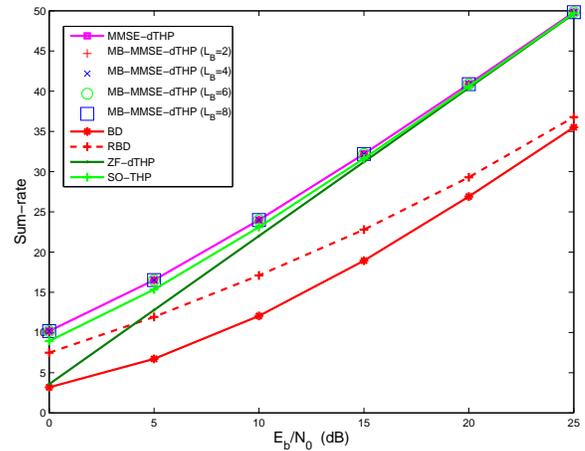}
      }
      \caption{Sum-rate performance of THP.}\label{THP_Sumrate}
    \end{figure}

Fig. \ref {cTHP_Sumrate} and Fig. \ref {dTHP_Sumrate} display the sum-rate performance of the proposed MB-MMSE-cTHP and MB-MMSE-dTHP algorithms, respectively.
From Fig. \ref {cTHP_Sumrate} , we can find that the sum rates of MB-MMSE-cTHP is improved with the increase of $L_B$ as we revealed in (59). When $L_B$ is increased to $4$, it can achieve almost the same sum-rate performance as with $8$ branches. The SO-THP in \cite{Veljko2} has shown a better sum-rate performance than MB-MMSE-cTHP algorithms for high values of $E_b/N_0$.
For MB-MMSE-dTHP, however, they share almost the same sum-rate performance with different branches. This phenomenon confirms the analysis developed in Section IV.  Another interesting phenomenon can be observed by comparing these two figures is that the sum-rate performance of MB-MMSE-cTHP is better than MB-MMSE-dTHP at low values of $E_b/N_0$, while MB-MMSE-dTHP offers a very good performance at high values of $E_b/N_0$.
%\begin{figure}[htp]
%\begin{center}
%\def\epsfsize#1#2{0.6\columnwidth}
%\epsfbox{MB_MMSE_cTHP_8X8_16QAM_Sumrate.eps} \vspace{-1.25em}
%\caption{\footnotesize Sum-rate performance of cTHP.} \label{cTHP_Sumrate}
%\end{center}
%\end{figure}

%\begin{figure}[htp]
%\begin{center}
%\def\epsfsize#1#2{0.6\columnwidth}
%\epsfbox{MB_MMSE_dTHP_8X8_16QAM_Sumrate.eps} \vspace{-1.25em}
%\caption{\footnotesize Sum-rate performance of dTHP.} \label{dTHP_Sumrate}
%\end{center}
%\end{figure}
\subsection{Correlated Channel State Information Scenario}
Here, we study the impact of correlated channels on the performance of the proposed and existing algorithms.
A correlated channel matrix can be obtained using the Kronecker model \cite{Paulraj01}
\begin{align}
\boldsymbol H_c=\boldsymbol R_r^{1\over 2}\boldsymbol H\boldsymbol R_t^{1\over 2}.
\end{align}
For the case of an urban wireless environment, the UE is always surrounded by rich scattering objects and the channel is most likely to be modeled by an independent Rayleigh fading channel at the receive side \cite{Keke03}. Hence, we assume $\boldsymbol R_r=\boldsymbol I_{N_r}$, and we have
\begin{align}
\boldsymbol H_c=\boldsymbol H\boldsymbol R_t^{1\over 2}.
\end{align}
To study the effect of antenna correlations, random realizations of correlated channels are generated according to the exponential correlation model \cite{Sergey} such that the elements of $\boldsymbol R_t$ are given by
\begin{align}
r_{i,j}=\left\{\begin{array}{ll} r^{j-i},& i\leq j\\ r_{j,i}^*,& i>j \end{array} \right.,|r|\leq 1,
\end{align}
where $r$ is the correlation coefficient between any two neighboring antennas. This correlation model is suitable for our study since, in practice, the correlation between neighboring channels is higher than that between distant channels. In the following Fig. \ref{Channel Corr}, we examine the performance of the proposed MB-MMSE-THP algorithms with $|r|=0.5$. The simulation results show that with the spatial correlation, the proposed MB-MMSE-THP algorithms still offer a better performance compared to their conventional counterparts and the MMSE-cTHP is more sensitive to the spatial correlation.
%\begin{figure}[htp]
%\begin{center}
%\begin{subfigure}[a]
%% \def\epsfsize#1#2{0.5\columnwidth}
%\epsfbox{MB_MMSE_THP_8X8_16QAM_BER_Scr.eps} % \vspace{-1.25em}
%\caption{\footnotesize BER with spatial correlation, 16-QAM.} \label{Fig. 11.}
%\end{subfigure}
%~
%\begin{subfigure}[b]
%% \def\epsfsize#1#2{0.5\columnwidth}
%\epsfbox{MB_MMSE_THP_8X8_16QAM_Sumrate_Scr.eps} % \vspace{-1.25em}
%\caption{\footnotesize Sum-rate with spatial correlation.} \label{Fig. 12.}
%\end{subfigure}
%\end{center}
%\end{figure}

%\begin{figure}
%        \centering
%        \begin{subfigure}[a]{0.3\textwidth}
%                \centering
%                \includegraphics[width=\textwidth]{MB_MMSE_THP_8X8_16QAM_BER_Scr}
%                \caption{BER with spatial correlation.}
%                \label{Fig. 11.}
%        \end{subfigure}%
%        ~
%        \begin{subfigure}[b]{0.3\textwidth}
%                \centering
%                \includegraphics[width=\textwidth]{MB_MMSE_THP_8X8_16QAM_Sumrate_Scr}
%                \caption{Sum-rate with spatial correlation.}
%                \label{Fig. 12.}
%        \end{subfigure}
%        %\caption{Pictures of animals}\label{fig:animals}
%\end{figure}
    \begin{figure}[!ht]
      \centering
      \subfloat[BER with spatial correlation, $(2,2,2,2)\times 8$ MIMO, 16-QAM.\label{BER_Corr}]{
        \includegraphics[scale=0.41]{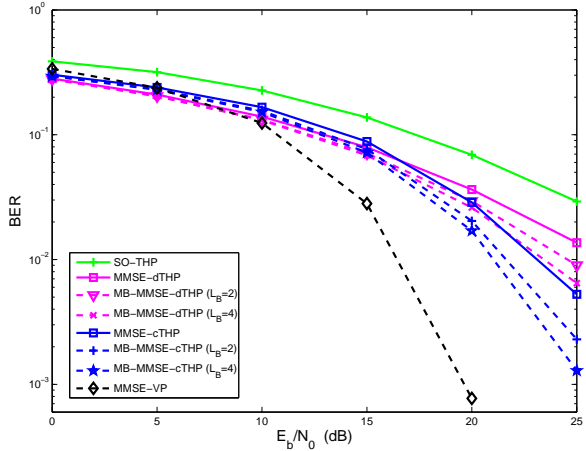}
      }
      \hspace{-10mm}
      \subfloat[Sum-rate with spatial correlation, $(2,2,2,2)\times 8$ MIMO, 16-QAM.\label{Sumrate_Corr}]{
        \includegraphics[scale=0.41]{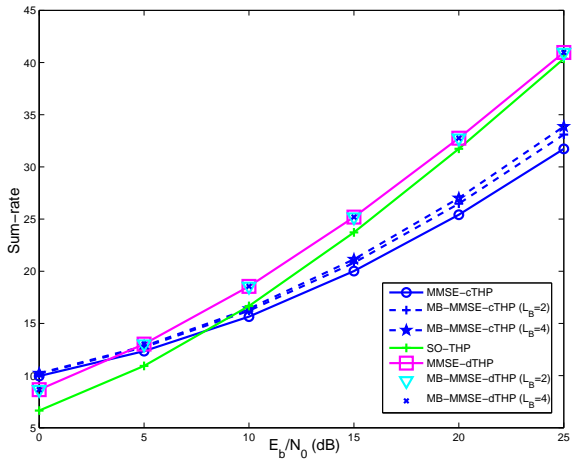}
      }
      \caption{Performance with correlated channel.}\label{Channel Corr}
    \end{figure}
%\begin{figure}[htp]
%\begin{center}
%\def\epsfsize#1#2{0.6\columnwidth}
%\epsfbox{MB_MMSE_THP_8X8_16QAM_Sumrate_Scr.eps} \vspace{-1.25em}
%\caption{\footnotesize Sum-rate with spatial correlation.} \label{Fig. 12.}
%\end{center}
%\end{figure}
\subsection{The impact of imperfect channels}
For the precoding techniques to work, CSI is required at the transmit side. This is natural for time-division duplex (TDD) systems because the uplink and downlink share the same frequency band. For frequency-division duplex (FDD) systems, however, the CSI needs to be estimated at the receiver and fed back to the transmitter. Assuming perfect CSI is impractical due to the often inaccurate channel estimation and the CSI feedback errors, we need to evaluate the impact of imperfect CSI on the performance of precoders. The channel errors can be modeled as a complex random Gaussian noise matrix $\boldsymbol E$ with i.i.d. entries of zero mean and variance $\sigma_e^2$. The imperfect channel matrix $\boldsymbol H_e$ is defined as \cite{Windpassinger}
\begin{align}
\boldsymbol H_e=\boldsymbol H+\boldsymbol E.
\end{align}

\begin{figure}[htp]
\begin{center}
\def\epsfsize#1#2{1.0\columnwidth}
\epsfbox{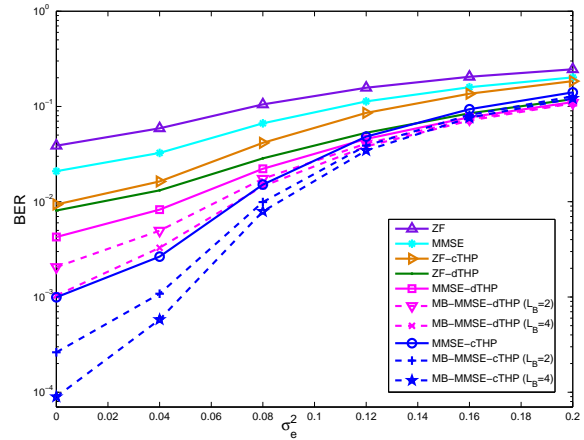} \vspace{-1.0em} \caption{\footnotesize BER as a
function of the variance of CSI error $\sigma_e^2$ for
$E_b/N_0=20~\rm dB$, $(2,2,2,2)\times 8$ MIMO, 16-QAM.}
\label{THP_ICH}
\end{center}
\end{figure}

Fig. \ref{THP_ICH} illustrates the BER performance of the above precoding algorithms with imperfect CSI at $E_b/N_0=20~\rm dB$. The BER performance gets worse for all the precoding algorithms with the increase of $\sigma_e^2$. The performance advantage of the proposed MB-MMSE-THP algorithms are not changed at low values of $\sigma_e^2$, while it degrades faster for higher values of $\sigma_e^2$ due to the nonlinearity of the algorithms.

It is worth noting that MMSE-cTHP loses its BER performance advantage to MMSE-dTHP for channel errors with a variance larger than $\sigma_e^2=0.14$ as shown in Fig. \ref{THP_ICH}, which illustrates that the cTHP structure is more sensitive than the dTHP structure to imperfect channels. Therefore, more feedback bits are needed by cTHP than dTHP in realistic systems. A robust optimization of THP based on the mean-squared-error (MSE) has been developed in \cite{Dietrich} to alleviate the impact of CSI errors. We leave a robust optimization under the MB-cTHP and MB-dTHP framework as a future extension.

Although less feedback information is required for dTHP in practice, the corresponding scaling matrix needs
to be transmitted to each distributed receiver, which requires an extra control overhead or additional feedforward information. Since the feedback issue is not the main focus of this work, we leave it for further research.

For MB-MMSE-cTHP and MB-MMSE-dTHP, we have found that
\begin {itemize}
\item{a better BER performance is obtained by MB-MMSE-cTHP compared to MB-MMSE-dTHP.}

\item{MB-MMSE-dTHP can lead to higher system sum rates and more flexible sub-channel management. The sum-rate performance of MB-MMSE-cTHP is not as good as MB-MMSE-dTHP but it can be improved with an increase of the number of branches.}

\item{MB-MMSE-cTHP is more sensitive to imperfect CSI than MB-MMSE-dTHP but a simplified receiver
structure for MB-MMSE-cTHP has been achieved due to the fact that
the decoding matrix $\boldsymbol G_k$ is not required at each
receiver in MB-MMSE-cTHP.}
\end {itemize}

%\begin{figure}[htp]
%\begin{center}
%\def\epsfsize#1#2{1.0\columnwidth}
%\epsfbox{MB_THP_Sumrate_ICH.eps} \vspace{-1.25em}
%\caption{\footnotesize Sum-rate with $\sigma_e^2$ for a fixed $E_b/N_0$.} \label{Fig. 14.}
%\end{center}
%\end{figure}
\section{\revnss{conclusions}}

%{\color {red}The joint successive detection cannot be implemented for the downlink of MU-MIMO systems due to the distributed users, however, we can equivalently implement THP at the transmit side}.
In this paper, MB-MMSE-cTHP and MB-MMSE-dTHP algorithms have been
proposed for MU-MIMO systems with multiple receive antennas. The
proposed MB-MMSE-THP algorithms exploit the degrees of freedom for
transmission by constructing a list of branches, which results in
extra transmit diversity gains. Moreover, the required computational
complexity is still reasonable since the filters of MB-MMSE-THP are
derived based on an LQ decomposition. A comprehensive performance
analysis has been carried out and a wide range of comparisons have
been conducted with existing precoding algorithms, including the BD,
RBD, THP, SO-THP, VP algorithms. Simulation results have illustrated
that a considerable improvement is achieved with only $2$ or $4$
branches, which reveals the value of the proposed MB-MMSE-THP
algorithms for practical applications. \revnss{Since a set of
parallel channels is obtained  with the application of the MB-THP
algorithms, power loading schemes can be employed to optimize the
power used over the channels.}
%\cite{Amico}

% with a reasonable computational complexity increase
%\section{conclusion}
%In this paper, a MB-THP algorithm has been proposed for the MIMO system and a significant BER gains are offered with a reasonable complexity.


\begin{thebibliography}{60}
{\footnotesize
%\bibitem{Alexiou}
%A. Alexiou and M. Haardt, "Smart antenna technologies for future wireless systems: Trends and challenges," IEEE Communications Magazine, vol. 42, issue:9, pp. 90 - 97, Sept. 2004.

%\bibitem{Spencer}
%Q. H. Spencer, C. B. Peel, A. L. Swindlehurst, and M. Haardt, "An introduction to the multi-user MIMO downlink," \textit{IEEE Communications Magazine}, pp. 60-67, Oct. 2004.

\bibitem{LTEA}
L. Liu, R. Chen, S. Geirhofer, K. Sayana, Z. Shi and Y. Zhou, "Downlink MIMO in LTE-Advanced: SU-MIMO vs. MU-MIMO," \textit{IEEE Commun. Mag.}, vol. 50, issue: 2, pp. 140-147, Feb. 2012.

\bibitem{Tse}
D. Tse and P. Viswanath, \textit{Fundamentals of wireless communications}. Cambridge University Press, 2005.

\bibitem{Windpassinger}
C. Windpassinger, "Detection and precoding for multiple input multiple output channels," Ph.D. dissertation, University Erlangen-Nurnberg, Germany, 2004.

\bibitem{Peel}
C. Peel, B. M. Hochwald and A. Swindlehurst, "A vector-perturbation technique for near capacity multiantenna multiuser communication - Part I: channel inversion and regularization," \textit{IEEE Trans. Commun.}, vol. 52, no. 1, pp. 195-202, Jan. 2005.

\bibitem{Michael}
M. Joham, W. Utschick and J. Nossek, "Linear transmit processing
in MIMO communications systems," \textit{IEEE Trans. Signal Process.}, vol. 53  no. 8, pp. 2700-2712, Aug. 2005.

\bibitem{Spencer}
Q. Spencer, A. Swindlehurst and M. Haardt, "Zero-forcing
methods for downlink spatial multiplexing in multiuser MIMO
channels," \textit{IEEE Trans. Signal Process.}, vol. 52, no. 2, pp. 461-471, Feb. 2004.

\bibitem{Choi}
L. Choi and R. Murch, "A transmit preprocessing technique for
multiuser MIMO systems using a decomposition approach," \textit{IEEE Trans. Wireless Commun.}, vol. 3, no. 1, pp. 20-24, Jan. 2004.

\bibitem{Veljko}
V. Stankovic and M. Haardt, "Generalized design of multi-user MIMO precoding matrices,"
\textit{IEEE Trans. Wireless Commun.}, vol. 7, no. 3, pp. 953-961, Mar. 2008.

\bibitem{Hochwald}
B. Hochwald, C. Peel and A. Swindlehurst, "A vector-perturbation technique for near capacity multiantenna multiuser communication - Part II: Perturbation," \textit{IEEE Trans. Commun.}, vol. 53, no. 3, Mar. 2005.% pp. 537 - 544, Mar. 2005.

\bibitem{Costa}
M. Costa, "Writing on dirty paper," \textit{IEEE Trans. Inform. Theory}, vol. 29, no. 3, pp. 439-441, May 1983.

\bibitem{Khina}
A. Khina and U. Erez, "On the Robustness of Dirty Paper Coding," \textit{IEEE Trans. Commun.}, vol. 58, no. 5, May 2010.

\bibitem{Tomlinson}
M. Tomlinson, "New automatic equaliser employing modulo arithmetic," \textit{Electronic Letters}, vol. 7, Mar. 1971.

\bibitem{Harashima}
H. Harashima and H. Miyakawa, "Matched-transmission technique for channels with intersymbol interference," \textit{IEEE Trans. Commun.}, vol. 20, Aug. 1972.

\bibitem{Fischer}
R. Fischer, C. Windpassinger, A. Lampe and J. Huber, "Space-time transmission using Tomlinson-Harashima precoding," in \textit{Proc. ITG Conf. on Source and Channel Coding (SCC)}, Berlin, Jan. 2002, pp. 139-147.

\bibitem{WY}
W. Y, D. Varodayan and J. Cioffi, "Trellis and convolutional precoding for transmitter based interference presubstration," \textit{IEEE Trans. Commun.}, pp. 1220-1230, Jul. 2005.

\bibitem{Erez}
U. Erez, S. Shamai and R. Zamir, "Capacity and lattice strategies for cancelling known interference," \textit{IEEE Trans. Inf. Theory}, pp. 3820-3833, Nov. 2005.

\bibitem{Christoph}
C. Windpassinger, R. Fischer, T. Vencel and J. Huber, "Precoding in
multiantenna and multiuser communications," \textit{IEEE Trans.
Wireless Commun.}, vol. 3, no. 4, Jul. 2004.

\bibitem{Windpassinger02}
C. Windpassinger, T. Vencel, and R. Fischer, "Precoding and loading
for BLAST-like systems," in \textit{Proc. IEEE Int. Conf. on Commun. (ICC)}, Anchorage, Alaska, USA, May 2003.

\bibitem{Joham01}
M. Joham, J. Brehmer, and W. Utschick, "MMSE approaches to multiuser
spatio-temporal Tomlinson-Harashima precoding," in \textit{Proc. 5th
ITG Conf. Source and Channel Coding (SCC)}, Germany, Jan. 2004.

\bibitem{Joham02}
M. Joham and W. Utschick, "Ordered spatial Tomlinson-Harashima precoding," in \textit{Smart Antennas-State-of-the-Art}, ser. EURASIP Book Series on Signal Processing and Communications. New York: Hindawi Publishing Corporation, 2005.

\bibitem{Wubben}
D.  W\"ubben, J. Rinas, R. B\"ohnke, V. K\"uhn and K. Kammeyer "Efficient algorithm for detecting layered space-time codes," in \textit{Proc. ITG Conf.
on Source and Channel Coding (SCC)}, Berlin, Germasy, Jan. 2002, pp. 399-405.

\bibitem{Wubben02}
D.  W\"ubben, R. B\"ohnke, V. K\"uhn and K. Kammeyer, "MMSE
extension of V-BLAST based on sorted QR decomposition," in
\textit{Proc. IEEE Vehicular Technology Conf. (VTC) Fall}, Orlando,
Florida, USA, Oct. 2003.

\bibitem{Jia01}
J. Liu and W. Krzymie\'n, "Improved Tomlinson-Harashima precoding
for the downlink of multi-user MIMO systems," \textit{Canadian
Journal of Electrical and Computer Engineering}, vol. 32, Summer
2007.

\bibitem {Habendorf}
R. Habendorf and G. Fettweis, "On ordering optimization for MIMO
systems with decentralized receivers," in \textit{Proc. IEEE
Vehicular Technology Conf. (VTC) Spring}, Melbourne, Australia, May
2006, pp. 1844-1848.

\bibitem{Jia02}
J. Liu and W. Krzymie\'n, "A Novel nonlinear joint
transmitter-receiver processing algorithm for the downlink of
multiuser MIMO systems," \textit{IEEE Trans. Veh. Technol.}, vol.
57, no. 4, Jul. 2008.

\bibitem{Fung}
C. Fung, W. Yu and T. Lim, "Precoding for the multiantenna downlink:
multiuser SNR gap and optimal user ordering," \textit{IEEE Trans.
Commun.}, vol. 55, no. 1, Jan. 2007.

\bibitem{Dao}
N. D\`ao and Y. Sun, "User-selection algorithms for multiuser precoding," \textit{IEEE Trans. Veh. Technol.}, vol. 59, no. 7, Sep. 2010.

\bibitem{Veljko2}
V. Stankovic and M. Haardt, "Successive Optimization Tomilinson-Harashima Precoding (SO-THP) for Multi-user MIMO systems," in \textit{Proc. IEEE Int. Conf. on Acoustics, Speech, and Signal Processing (ICASSP)}, Philadelphia, PA, Mar. 2005, pp. 1117-1120.

\bibitem{Keke}
K. Zu, R. C. de Lamare and M. Haardt, "Multi-Branch
Tomlinson-Harashima precoding for single-user MIMO systems," in
\textit{Proc. ITG/IEEE Workshop on Smart Antennas}, Dresden,
Germany, Mar. 2012.

\bibitem{MH}
M. Huang, S. Zhou and J. Wang, "Analysis of Tomlinson-Harashima precoding in multiuser MIMO systems with imperfect channel state information," \textit{IEEE Trans. Veh. Technol.}, vol. 57, no. 5, Sep. 2008.

\bibitem{Shiu}
D. Shiu and J. Kahn, "Layered Space-Time Codes for Wireless Communications using Multiple Transmit Antennas," in \textit{Proc. IEEE Int. Conf. on Communications (ICC'99)}, Vancouver, B.C., Canada, Jun. 1999.

\bibitem{Abay}
U. Abay and R. Fischer, "Comparison of generalized Tomlinson-Harashima precoding strategies for the broadcast channel," in \textit{Proc. ITG/IEEE Workshop on Smart Antennas}, Aachen, Germany, Feb. 2011.

\bibitem{Dietrich}
F. Dietrich, P. Breun and W. Utschick, "Robust Tomlinson-Harashima
precoding for the wireless broadcast channel," in \textit{IEEE
Trans. Signal Process.}, vol. 55, no. 2, Feb. 2007.

\bibitem{Rodrigo}
R. C. de Lamare and R. Sampaio-Neto, "Minimum mean squared error
iterative successive parallel arbitrated decision feedback detectors
for DS-CDMA systems," \textit{IEEE Trans. Commun.}, May 2008.

\bibitem{Rui}
R. Fa and R. C. de Lamare, "Multi-branch successive interference cancellation for MIMO spatial multiplexing systems: design, analysis and adaptive implementation," \textit{IET Commun.}, May 2010.

\bibitem{Rodrigo02}
Y. Cai, R. C. de Lamare, R. Fa, "Switched interleaving techniques with limited feedback for interference mitigation in DS-CDMA systems," \textit{IEEE Trans. Commun.}, vol. 59, no. 7, Jul. 2011.

\bibitem{mbdf}
R. C. de Lamare, ``Adaptive and Iterative Multi-Branch MMSE Decision
Feedback Detection Algorithms for Multi-Antenna Systems," \emph{IEEE
Trans. Wireless Commun.}, vol.12, no.10, pp.5294-5308, October 2013.

\bibitem{MatrixTheo}
G. Golub and C. Van Loan, \textit{Matrix Computations}. The Johns Hopkins University Press, 1996.

\bibitem{Zu}
K. Zu and R. C. de Lamare, "Low-complexity lattice reduction-aided regularized block diagonalization for MU-MIMO systems," \textit{IEEE Commun. Lett.}, 2012.

\bibitem{Zu02}
K. Zu, R. C. de Lamare and M. Haardt, "Generalized design of
low-complexity block diagonalization type precoding algorithms for
multiuser MIMO systems," \textit{IEEE Trans. Commun.}, vol. 61, no.
10, Oct. 2013.

\bibitem{Kusume}
K. Kusume, M. Joham, W. Utschick and G. Bauch, "Cholesky
factorization with symmetric permutation applied to detecting and
precoding spatially multiplexed data streams," \textit{IEEE Trans.
Signal Process.}, vol. 55, no. 6, Jun. 2007.

\bibitem{MMSE-VP}
D. Schmidt, M. Joham and W. Utschick, "Minimum mean square error vector precoding," \textit{European Trans. on Telecommun.}, vol. 19, no. 3, 2008.

\bibitem{SD}
B. Hassibi and H. Vikalo, "On the sphere decoding algorithm: Part I the expected complexity," \textit{IEEE Trans. Signal Process.}, vol. 53, no. 8, pp. 2806-2818, Aug. 2005.

\bibitem{Keke02}
K. Zu and R. C. de Lamare, "Pre-sorted multiple-branch successive interference cancelation detection for high-dimensional MIMO systems," in \textit{Proc. ITG/IEEE Workshop on Smart Antennas}, Dresden, Germany, Mar. 2012.

\bibitem{Paulraj01}
A. Paulraj, R. Nabar and D. Gore, \textit{Introduction to space-time wireless communications}. Cambridge University Press, 2003.

\bibitem{Keke03}
K. Zu, R. C. de Lamare and M. Haardt, "Lattice reduction-aided regularized block diagonalization for multiuser MIMO systems", \textit{IEEE Wireless Commun. and Networking Conf. (WCNC)}, Paris, France, Apr. 2012.

\bibitem{Sergey}
S. Loyka, "Channel capacity of MIMO architecture using the exponential correlation matrix," \textit{IEEE Commun. Lett.}, vol. 5, no. 9, pp. 369-371, Sep. 2001.

%\bibitem{Amico}
%A. Amico and M. Morelli, "Joint Tx-Rx MMSE Design for MIMO Multicarrier Systems with Tomlinson-Harashima Precoding," \textit{IEEE Trans. Commun.}, vol. 7, no. 8, pp. 3118-3127, Aug. 2008.
}
\end{thebibliography}
\end{document}